\documentclass{aa}

\usepackage{graphicx}
\usepackage{txfonts}
\usepackage{hyperref}
\usepackage[capitalise]{cleveref}
\usepackage{tabularx}
\usepackage{multicol}
\usepackage{stfloats}
\usepackage{multirow}
\usepackage{booktabs}
\usepackage{subcaption}
\usepackage{threeparttable}
\usepackage{placeins}

\usepackage{afterpage}
\usepackage{todonotes}
\usepackage{natbib}
\bibpunct{(}{)}{;}{a}{}{,} 
\usepackage{verbatim}
\renewcommand{\rm}[1]{\mathrm{#1}}

\begin{document}

   \title{Outflowing shocked gas dominates the NIR H$_2$ emission from the dual AGN NGC~6240}
    \author{J. Carlsen \inst{1}
    \and C. Cicone \inst{1}\thanks{Corresponding author: claudia.cicone@astro.uio.no}
    \and B. Hagedorn \inst{1}
    \and K. Rubinur \inst{1}
    \and P. Andreani \inst{2}
    \and K. Dasyra \inst{3}
    \and P. Severgnini \inst{4}
    \and C. Vignali \inst{5,6}
    \and R. Morganti \inst{7,8}
    \and T. Oosterloo \inst{7,8}
    \and A. Lasrado \inst{1}
    \and E. Lopez-Rodriguez \inst{9}
    \and S. Shen \inst{1}
    }
    \institute{\inst{1}Institute of Theoretical Astrophysics, University of Oslo, P.O. Box 1029, Blindern, 0315 Oslo, Norway \\ 
    \inst{2} European Southern Observatory (ESO), Karl-Schwarzschild-Strasse 2, D-85748 Garching, Germany \\ 
    \inst{3} Section of Astrophysics, Astronomy \& Mechanics, Department of Physics, National and Kapodistrian University of Athens, University Campus Zografos, GR 15784 Athens, Greece \\ 
    \inst{4} INAF - Osservatorio Astronomico di Brera, Via Brera 28, 20121 Milano, Italy \\ 
    \inst{5} Dipartimento di Fisica e Astronomia Augusto Righi, Università degli Studi di Bologna, via Gobetti 93/2, 40129 Bologna, Italy \\ 
    \inst{6} INAF – Osservatorio di Astrofisica e Scienza dello Spazio di Bologna, Via Gobetti 101, 40129 Bologna, Italy \\ 
    \inst{7} ASTRON, The Netherlands Institute for Radio Astronomy, Oude Hoogeveensedijk 4, 7991 PD, Dwingeloo, The Netherlands \\ 
    \inst{8} Kapteyn Astronomical Institute, P.O. Box 800, 9700 AV Groningen, The Netherlands \\ 
    \inst{9} Department of Physics \& Astronomy, University of South Carolina, Columbia, SC 29208, USA \\ 
       }

   \date{}
 
  \abstract{We present a multi-line study of the kinematics of the molecular and ionised gas phases in the central $\sim~2~$kpc of the luminous infrared galaxy and dual AGN NGC~6240, based on archival JWST/NIRSpec and ALMA observations. Since a primary goal of our analysis is to study outflows, we devised a new spectral-line fitting approach to de-blend rotating and non-rotating gas components in the observed near-infrared (NIR) emission lines. Our method is more physically motivated than previous approaches and better tailored to the extreme feedback mechanisms at work in NGC~6240.
  We find that $\sim65$\% of the Pa$\alpha$, H$_2$, and [FeII] line fluxes within the NIRSpec field of view arise from gas components that are kinematically decoupled from the stars. In particular, the NIR H$_2$ lines show the most deviation from the stars, with peak emission between the two rotating stellar structures. The polycyclic aromatic hydrocarbon (PAH) emission feature at 3.3~$\mu$m does not follow the NIR H$_2$ morphology, indicating that the latter does not trace photon-dominated regions. In the non-rotating gas components, we identify a bi-conical wind launched from the northern AGN, expanding along the minor axis of stellar rotation. This wind is dominated by ionised gas and, although it entrains some H$_2$ gas, it does not show a H$_2$/PAH enhancement, suggesting either high UV irradiation or expansion along a relatively gas-free path. Furthermore, we find bright non-rotating gas emission between the two AGN and around the southern AGN, which we interpret as due to an outflow launched from the southern nucleus, coinciding with the massive molecular outflow previously studied in cold (sub-)millimeter tracers. The strong H$_2$/PAH enhancement measured in this region, coextensive with high velocity redshifted gas ($v\sim900$~km~s$^{-1}$), suggests that the shocks responsible for the high H$_2$/PAH ratios are due to the outflow rather than to the collision of media during the merger. Our results show that the bulk of the NIR line emission in NGC~6240 is decoupled from the stars, and that most of the warm H$_2$ is shock-excited and embedded in a powerful outflow, where it coexists with colder molecular gas.}

  
\titlerunning{A massive outflow of shocked gas in NGC~6240}
\authorrunning{J. Carlsen et al.}
\maketitle
%
 
\section{Introduction}\label{sec:Intro}

Major mergers play an important role in galaxy evolution by providing fuel for starbursts and accreting supermassive black holes (SMBHs) \citep{mihos_gasdynamics_1996}. Quasar activity and the creation of elliptical galaxies are also believed to be caused by mergers \citep{Engel+10}. In the local Universe ($z\leq0.2$), ultra-luminous infrared galaxies (ULIRGs), defined by their high infrared (IR) luminosity $L_\rm{IR (8-1000\mu m)}\geq10^{12}\,L_\odot$ \citep{Sanders+Mirabel96}, correspond to gas- and dust-rich major galaxy mergers \citep{lonsdale_ultraluminous_2006}. The combination of a starburst and/or an active galactic nucleus (AGN) in a dust-enshrouded environment is thought to be responsible for their extreme IR luminosity  \citep{sanders_ultraluminous_1988,veilleux_spitzer_2009}. Studying the local ULIRG population is crucial to gain insights into such a short-lived, but potentially disruptive, phase of galaxy evolution. During this phase, galaxies experience intense merger-driven starburst and quasar activity and, consequentially, their interstellar medium (ISM) is subject to strong radiative and mechanical feedback mechanisms \citep{perez-torres_star_2021,u_role_2022}.
A clear manifestation of such feedback is galactic-scale (e.g., kpc-scale) outflows.  
Indeed, the energy released by AGN and starbursts drives massive outflows, which can reduce the star formation rate (SFR) and accretion onto the central SMBH. 
In ULIRGs, outflows are powerful, multi-phase, and dominated in mass by cold molecular gas \citep{Sturm+11,Aalto+12,Spoon+13,Veilleux+13,Cicone+14,Cicone+20,Veilleux+20,Fluetsch+21}. Massive molecular outflows have an incidence of $>70\%$ in ULIRGs \citep{Veilleux+13,Lamperti+22}.

It is generally accepted that mergers of two massive galaxies lead to the formation of systems of SMBH pairs, initially at kpc and sub-kpc separations (dual SMBHs) and, as the galaxies coalesce, reaching down to pc and sub-pc separations (binary SMBHs) \citep{begelman+1980, merritt+2005}. Through the accretion of the surrounding gas, one or both SMBHs may shine as AGN \citep{Dotti+07}, and so be detectable via their electromagnetic signals. When both SMBHs in a pair are simultaneously active, we refer to the system as a dual or binary AGN, depending on the physical separation of the pair members \citep{DeRosa+19, rubinur+2019}. The search for dual and binary AGN has recently gained more consideration,
driven by the potential of current and future gravitational wave (GW) experiments \citep{Colpi+24_LISA, EPTA2023_paper} of detecting signals emitted by these systems. Being the recent product of major mergers, local ULIRGs are expected to harbour dual SMBHs. However, confirming their presence can be particularly difficult due to several factors: the higher nuclear dust obscuration of these sources \citep{Ricci+21}, the expected low percentage of simultaneously accreting SMBHs detectable as dual AGN \citep{VanWassenhove+12}, and the scarcity of multiwavelength data coverage with sufficient angular resolution to distinguish the pair members and confirm the AGN nature. As a result, only a few ULIRGs host confirmed dual AGN, and the target of our study, NGC~6240, is the best studied case.

NGC~6240 results from the collision of two gas-rich galaxies of similar mass \citep{Fosbury+79}. The merger geometry tends towards coplanar/prograde, and the source is captured between first encounter and coalescence stage \citep{Engel+10}. Although formally classified as LIRG with $L_{IR(8-1000\mu m)}=10^{11.7}~L_{\odot}$ \citep{Liu+21}, NGC~6240 is often considered a {\it bona fide} ULIRG \citep{Genzel+98}. It hosts two confirmed AGN with a projected separation between $1.5''$ and 1.8$''$ (800 - 900 pc) at X-ray \citep{Komossa+03, nardini+13} and radio wavelengths \citep{Hagiwara+11}. The two AGN reside at the centres of two stellar rotating structures, identified by several authors \citep{Engel+10, Ilha+16, Kollatschny+20}, and most recently by \cite{ceci_jwstnirspec_2024} using the same JWST data used in this work. The sub-millimetre/millimetre (hereafter, (sub-)mm) continuum, tracing thermal cold dust emission, shows two distinct peaks near the AGN positions \citep{treister_molecular_2020, Scoville+15}.

The morphology and kinematics of the molecular gas in NGC~6240 are puzzling, and, to a large extent, inconsistent with expectations from current hydrodynamical simulations of galaxy mergers. 
Indeed, unlike ionised gas tracers, none of the molecular emission lines explored so far, e.g. several CO, [CI], HCN, CS, HCO$^+$ transitions at (sub-)mm wavelengths \citep{Tacconi+99, Iono+07, Scoville+15, Feruglio+13a, Cicone+18} and near-IR (NIR) H$_2$ ro-vibrational lines \citep{Herbst+90, vanderWerf+93} present two distinct surface brightness peaks, as it would be expected if the H$_2$ gas followed the dust and stellar distributions. Instead, all H$_2$ tracers in NGC~6240 hint at a high concentration of gas in the internuclear region between the two AGN. This structure is elongated towards the south, almost reaching the southern AGN in the (sub-)mm lines, with a clear peak near the southern AGN in the NIR H$_2$(1-0) S(1) line \citep{Max+05}. This line is exceptionally bright in NGC~6240 \citep{Goldader+95}.
\cite{Max+05} suggested the central molecular gas structure of NGC~6240 to be a bridge connecting the two nuclei, possibly tracing gas flowing from one nucleus to another (see also \citealt{treister_molecular_2020}). \cite{Tacconi+99} interpreted the bright CO-emitting nuclear reservoir as a rotating and highly turbulent molecular gas disc, but this hypothesis is inconsistent with the presence of two stellar discs around the nuclei and has been ruled out by high resolution data \citep{treister_molecular_2020, Johnstone+25}.

The molecular gas properties of NGC~6240 must be analysed within the framework of the powerful and disruptive feedback processes at work in this source. Ionised gas outflows are a clear manifestation of such feedback, and include bi-conical or conical winds, bubbles, \citep{muller-sanchez_two_2018, ceci_jwstnirspec_2024, Hermosa-Munoz+25}, and a superwind expanding into the circumgalactic medium detected in H$\alpha$ \citep{Heckman+90, Armus+90, Gerssen+04, Veilleux+03, yoshida_giant_2016} and soft X-ray emission \citep{nardini+13}. In addition, NGC~6240 hosts a massive outflow of molecular gas. The latter was
first suggested by high-velocity components of the $H_2$ 1-0 S(1) emission line \citep{vanderWerf+93, Ohyama+00, Ohyama+03}, and then unambiguously confirmed through highly blue-shifted ($v_{max}=-1200$~km~s$^{-1}$) OH absorption \citep{Veilleux+13} and high sensitivity, wide bandwidth CO line observations \citep{Feruglio+13a, Feruglio+13b}. Its properties are extreme even compared to other massive molecular outflows detected in local ULIRGs: it extends by more than 10~kpc along the east-west direction, entrains $>10^{10}$~M$_{\odot}$ of H$_2$ gas, i.e. $\sim60$~\% of the cold molecular gas reservoir of the system, and it dominates the cold gas kinematics in the central 1-2 kpc \citep{Cicone+18}.

Here we present a new analysis of the morphology and kinematics of molecular and ionised gas tracers in NGC~6240, primarily based on public archival JWST NIRSpec observations recently presented by \cite{ceci_jwstnirspec_2024} and supplemented with ALMA data from \cite{Cicone+18} and from the public archive. Differently from  \cite{ceci_jwstnirspec_2024}, we do not rely on an arbitrary decomposition of narrow and broad spectral line components to identify outflow features. We devise a more physically motivated approach that exploits the kinematic parameters derived from the stellar fit to disentangle disc and non-disc components in the gas tracers. As we discuss in the paper, this method allows us to discover significant additional H$_2$ components entrained in the outflow and to recognise a striking overlap between the cold and warm phase of the massive molecular outflow in NGC~6240, providing further clues to its origin.

Throughout this paper, the cosmology is the standard $\Lambda$CDM model with $H_0=67.8\,\rm{km\,s^{-1}Mpc^{-1}}$, $\Omega_\rm m=0.308$, and $\Omega_\Lambda=0.692$. At the distance of NGC~6240 with redshift $z=0.02448$ and luminosity distance $D_L=110.3$ Mpc, the physical scale is 0.509 kpc arcsec$^{-1}$.

\section{Data}
\subsection{Archival JWST NIRSpec observations}\label{sec:data_jwst}
\begin{figure}
    \centering
    \includegraphics[clip=true,trim=0.1cm 0.3cm 0.0cm 0.1cm, width=\columnwidth]{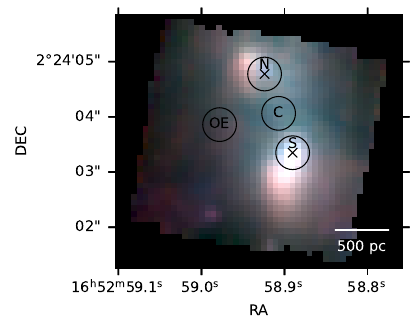}
    \caption{RGB composite map of NGC~6240 obtained with JWST NIRSpec, tracing the emission at $\sim1.6~\mu$m (blue), 2.9~$\mu$m (green), and 4.6~$\mu$m (red). Four representative apertures with $0.3''$ radius are shown on the map: the VLBI positions \citep{Hagiwara+11} of the two known AGN (N and S) at $\rm{R.A.=16^h52^m58^s.9241}$, $\rm{Dec.=02^\circ24'04''.776}$, and $\rm{R.A.=16^h52^m58^s.8902}$, $\rm{Dec.=02^\circ24'03''.350}$ respectively;
    a central position between the nuclei (C), at R.A.=16$^h$52$^m$58$^s$.907, Dec.=02$^{\circ}$24$'$4$''$.063, and an offset point to the east (OE), at R.A.=16$^h$52$^m$58$^s$.978, Dec.=02$^{\circ}$24$'$3$''$.859. The physical scale of the map (500 pc) is indicated by the white line.}
    \label{fig:rgb}
\end{figure}

Our analysis relies on the NIR integral field unit (IFU) observations 
of the nuclear region of NGC~6240, collected with the JWST/NIRSpec instrument as part of the 
Guaranteed Time Observations (GTO) program 1265 (PI: A. Alonso-Herrero), on July 17, 2023. 
The data reduction was done using the JWST pipeline v1.14.0 with CRDS context 1242. The point-spread function (PSF) of NIRSpec is $0.08''$ \citep{Jakobsen+22}. We gathered the combined science (stage 3) data products from the MAST web portal\footnote{\url{https://mast.stsci.edu/portal/Mashup/Clients/Mast/Portal.html}}.

Three high-resolution ($R\sim2700$) grating/filter configurations were used: G140H/F100LP covering the wavelength range 0.97-1.89 $\mu$m, G235H/F170LP covering 1.66-3.17 $\mu$m, and G395/F290LP covering 2.87-5.27 $\mu$m. We will refer to these configurations as the 1.0 $\mu$m band, 1.7 $\mu$m band, and 2.9 $\mu$m band, respectively. Each band has an effective integration time of 219 seconds. In \cref{fig:rgb}, we show a composite image of the NIRSpec field-of-view (FoV) ($5.1''\times4.5''$, i.e. 2.5~kpc $ \times$ 2.3~kpc), where the red, green, and blue channels correspond to wavelengths around 1.6, 2.9, and 4.6~$\mu$m, respectively. The spaxel size is $0.1''\times0.1''$. Four apertures with $0.3''$ radius are shown as well, covering the northern (N) and southern (S) nuclei (hereafter, N AGN and S AGN), a region offset to the east of the two nuclei (OE), and a central (C) region between them. The two AGN are marked using the Very-Long-Baseline Interferometry (VLBI) positions from \cite{Hagiwara+11}.

We perform an astrometric corrections to all three bands, following \cite{ceci_jwstnirspec_2024}, who analysed the same data. We assume that the $4.5~\mu$m continuum peak coincides with the S AGN, the brighter one. The correction offsets applied to the data are $\rm{\delta R.A.=-0.22''}$  and $\rm{\delta Dec.}=0.17''$. After the correction, both nuclei are consistent with the VLBI AGN positions.

As a result of the under-sampling of the PSF, NIRSpec IFU single-spaxel spectra close to bright point sources often display sinusoidal artifacts commonly referred to as wiggles. Although more pronounced in data cubes with spaxel size of $0.05''$, these artifacts also partially affect our data. More specifically, we detect wiggles over an approximately circular area centred on the S AGN with $r\sim0.2''-0.3''$ (see top-right panel of \cref{fig:ppxf-results}), but not in other portions of the FoV. We have tested the possible impact of the wiggles on the results of our spectral fitting procedure (described in Sect.~\ref{sec:results_spectralfit}) in the most severely affected spaxel at the S AGN position, and found no statistically significant effect. We have also repeated the whole analysis by masking the 15 spaxels that are affected by wiggles, which are all around the S AGN. The resulting line fits and flux maps remain unchanged within the uncertainties, and do not affect any of the conclusions presented in this work.
We note that all wiggle correction techniques in use in the literature (e.g. \citealt{perna_ga-nifs_2023})
rely on modelling the wiggles from the same scientific dataset under study. Such an empirical approach may not deliver perfect results (see \citealt{ceci_jwstnirspec_2024}). It is especially risky in our case, where extended emission from gas and stars overlaps with the point sources; thus, the only point-source spectrum available in our dataset near the S AGN, which should be used to model the wiggles, displays line wings of physical origin, due to known widespread outflows, as well as stellar absorption lines. For all these reasons, we decided not to perform a wiggle correction on our data. 

\subsection{Archival ALMA observations}
\subsubsection{Band 7 (870$\mu$m) continuum data}
We use the high-resolution data available in the ALMA archive as part of project ID 2013.1.00813.S (PI: Rangwala). In particular, we use the ALMA Science data model (Asdm) \texttt{uid://A002/Xa8df68/X101b}. The calibrated measurement set (MS) was kindly provided upon request by the ALMA European ALMA Regional Center. The observations were taken with 41 antennas of the ALMA 12m array on 27 August, 2015, with a precipitable water vapor (PWV) of 0.813 mm. The total observing time was 28 minutes, and the on-source time on NGC~6240 was a little over 3 minutes. The following sources were used as calibrators: J1733-1304 (Bandpass), Ceres (Amplitude, Flux), J651+0129 (Phase), J1649+0412 (Delay). 

From this dataset, we produce a map of the 870$\mu$m dust continuum in the nuclear region of NGC~6240, with an angular resolution comparable to that of JWST NIRSpec. 
The (sub-)mm spectrum of NGC~6240 is chemically rich, and most species are bright and so detectable by ALMA even with short exposures. Moreover, due to the massive molecular outflow (see Sect~\ref{sec:Intro}), most molecular transitions in this source show broad wings, so one has to worry about line contamination even in spectral channels that are distant from the central frequency of a line. We thus image and visually inspect each spectral window to select the line-free channels for continuum imaging. Besides the obvious contamination from the broad CO(3-2) line (at $\nu_{obs} = 337.5$~GHz) that occupies most of spw~2 and part of spw~3 of this dataset, we find a possible signature of contamination from other species such as HCO$^+$(4-3) and SiO(8-7). We finally conservatively select the line-free channels using the \texttt{CASA} task \texttt{split}, i.e.: $\nu_{obs}\in (349.5, 350.4)$~GHz in spw 0, $\nu_{obs}\in (350.3, 351.9)$~GHz in spw 1, and $\nu_{obs}\in (339.5, 340.2)$~GHz in spw 3. We then image the corresponding uv visibilities using the task \texttt{tclean} with \texttt{specmode = mfs}, using a hogbom deconvolver, with briggs weighting (robust = 0.5). We use a cleaning mask centred at RA=16h52m58.91s, Dec=02d24m03.89s with a radius of $1.3~''$. The cell size is $0.05''$. The resulting map has a synthesized beam of $0.315''\times 0.14''$ (PA=-65.7~deg). We apply a primary beam correction using the \texttt{impbcor} task. The final sensitivity of the map is 0.3~mJy/beam ($1\sigma$ RMS). We note that the same data are used by \cite{Fyhrie+21} and their Band~7 continuum map is consistent with the one produced by us.

\subsubsection{Band 8 [CI]$^3P_1-^3P_0$ line data}
The ALMA Band~8 data from project ID 2015.1.00717 (PI: Cicone) enable us to compare the morphology of the cold molecular outflow as imaged using the atomic carbon [CI]$^3P_1-^3P_0$ line (hereafter, [CI](1-0)) with that of the warm molecular hydrogen observed by JWST. We use the same ALMA maps as shown in \cite{Cicone+18} (panels c and d of Figure 1 in that paper), which were obtained by integrating the [CI](1-0) line emission respectively between (-650, -200)~km~s$^{-1}$ (blue wing) and (250, 800)~km~s$^{-1}$ (red wing). We refer to that paper for technical details about these observations as well as the analysis that led to the computation of the molecular outflow properties. The synthesized beam of the ALMA [CI](1-0) maps is $0.29''\times 0.24''$ (PA=113.2~$\deg$).

\subsubsection{Band 3 CO(1-0) line data}
To strengthen our comparison between the cold phase of the molecular gas and the warm phase traced by the NIR H$_2$ lines, we also use the ALMA Band~3 CO(1-0) line data from project ID 2015.1.00003.S (PI: Saito). The calibrated MS was provided by the ALMA European ALMA Regional Center. The Asdm is \texttt{uid://A002/Xb68dbd/X72f9}. The observations were taken with 37 antennas of the 12m ALMA array on 16 Aug 2016 with excellent weather conditions for Band~3 (PWV = 0.57mm). The total observing time was 28 minutes, and the on-source time on NGC~6240 was 10 minutes. The following sources were used as calibrators: J1550+0527 (bandpass), Titan (flux), J1651+0129 (Phase). We focus the analysis on spw~0, which covers the CO(1-0) line centered at $\nu_{obs}=112.52$~GHz. The native spectral resolution of the data is $\delta v_{res}=2.6$~km~s$^{-1}$. We use the task \texttt{uvcontsub} to fit the continuum emission with a first-order polynomial and subtract it from the uv visibilities. The frequency ranges selected to estimate the continuum level are $\nu_{obs}\in (111.6, 112.1)$~GHz and  
$\nu_{obs}\in (113, 113.4)$~GHz. 
We image the continuum-subtracted visibilities using \texttt{tclean}, \texttt{specmode=cube}, re-binning the data using a channel width of 5. We select the hogbom deconvolver and use Briggs weighting with robust = 0.5. The mask is defined using interactive cleaning to adjust it to the extent of the source, which varies for different channels. The cell size is $0.08''$, and the resulting synthesized beam is $0.55''\times 0.52''$, PA=-52.8~deg. The spectral bandwidth covers a velocity range of (-2521, 2462)~km~s$^{-1}$. The average line sensitivity is 0.6~mJy/beam per 13~km~s$^{-1}$ channel.

\section{Results}

\subsection{Spectral properties}

\begin{figure*}[t]
    \centering
    \includegraphics[clip=true,trim=0.1cm 0.3cm 0.3cm 0.2cm, width=\textwidth]{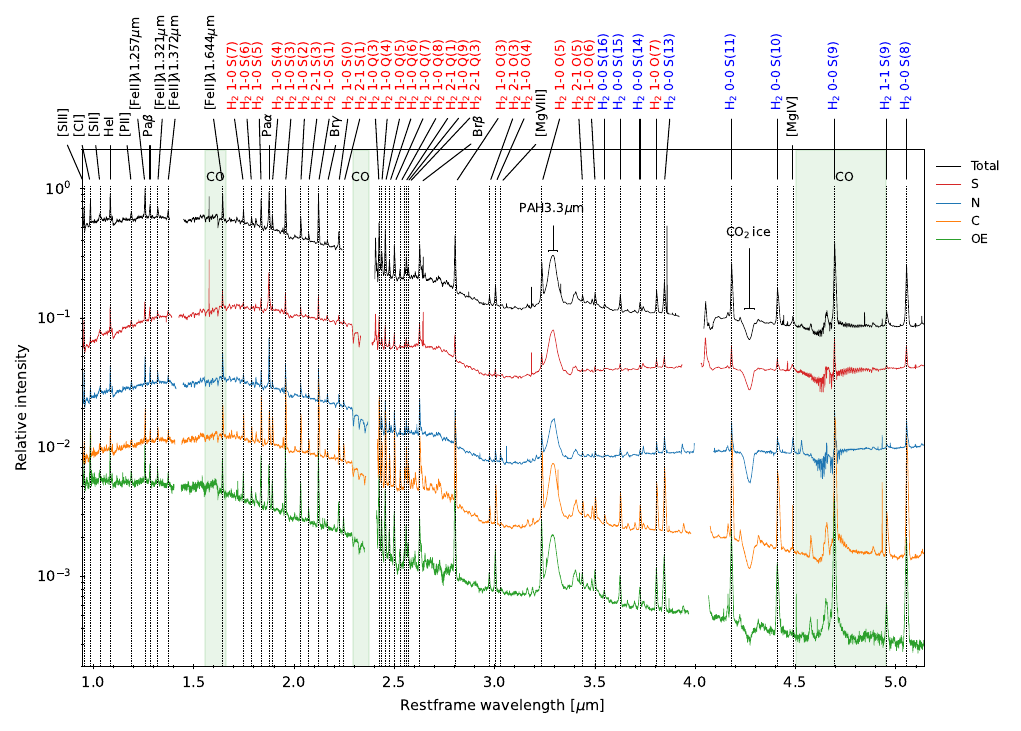}
    \caption{Relative intensity of spectra extracted using an aperture of 3 spaxels from the northern (N) and southern (S) nuclei, a centre (C) point between the AGNs, an offset (OE) point (see \cref{fig:rgb}), and the full FoV (Total). The spectra are joined from the three JWST/NIRSpec grating/filter compositions to cover all the wavelengths. CO bands are marked as shaded regions. Atomic lines are annotated in black, ro-vibrational H$_2$ transitions in red, and pure rotational H$_2$ transitions in blue.}
    \label{fig:gather}
\end{figure*}

We extract the NIR spectra from positions N, S, OE, and C (see \cref{fig:rgb}) using apertures with radius $0.3''$, and show them in \cref{fig:gather}, together with a total spectrum extracted from the full FoV. The flux densities have been normalized and scaled relative to the peak of the total spectrum for visual purposes and the spectra are shown in rest-frame wavelengths, using $z=0.02448$. The three gaps in the spectra correspond to the gaps between the NIRSpec detectors. We identify several emission lines, marked with vertical dashed lines, most of which are ro-vibrational and purely rotational lines of molecular hydrogen ($\rm{H_2}$, see \cref{tab:identified-lines}). The polycyclic aromatic hydrocarbon (PAH) emission feature at 3.3~$\mu$m is prominent in all spectra.
We mark three stellar CO absorption regions: (i) from CO(3-0) to CO(8-5) at $\lambda=$1.5528-1.6610 $\mu$m; (ii) from CO(2-0) to CO(3-1) at $\lambda=$2.2935-2.3739 $\mu$m; and (iii) from CO(1-0) R(22) to P(19) at $\lambda=$4.5010-4.9486 $\mu$m.
We have also identified the CO$_2$ bending-vibrational mode at $4.27\,\mu$m. 
The comparison between the spectra in \cref{fig:gather} shows: (i) significantly weaker ro-vibrational and rotational H$_2$ lines at both AGN positions compared to regions C and OE; (ii) a flat or even slightly rising continuum at $\lambda>3.5~\mu$m in the AGN spectra, whereas the opposite is true for the C and OE spectra; (iii) stronger Pa$\alpha$ and Br$\gamma$ emission in the N AGN compared to the other spectra; and (iv) weaker [FeII] emission in the S AGN than in the other regions.
Such spatial variations will be further explored in the analysis that follows, which will focus mainly on the kinematics and morphology of the following tracers: Pa$\alpha$, Pa$\beta$, H$_2$ 1-0 S(1), H$_2$ 0-0 S(8), [FeII]$\lambda1.257\mu$m, and [FeII]$\lambda1.644\mu$m, as well as PAH 3.3$\mu$m. A detailed NIR spectral analysis is beyond the scope of this paper, as it was already done by \cite{ceci_jwstnirspec_2024}. 

\subsection{Stellar kinematics}\label{sec:results_stellarkin}

\begin{figure*}[t]
    \centering
    \includegraphics[width=0.72\textwidth]{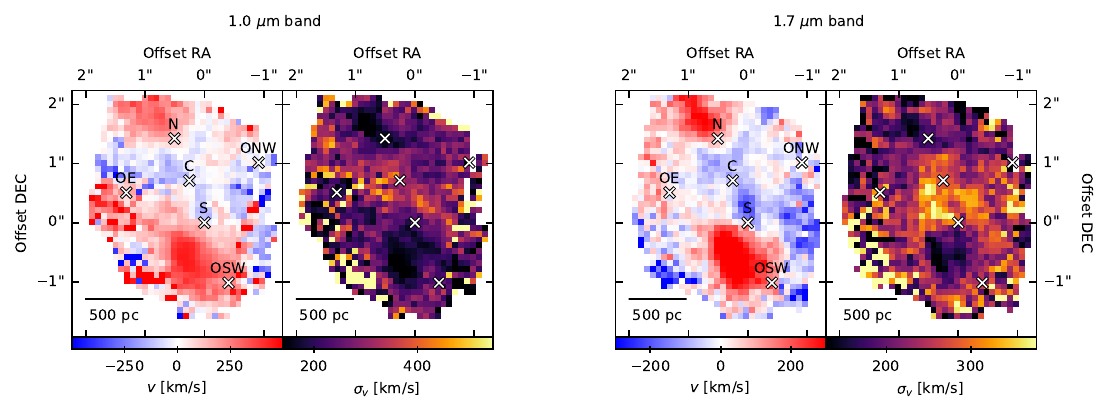}\quad
    \includegraphics[clip=true,trim=0.1cm 0.1cm 0.3cm 0.2cm, width=0.25\textwidth]{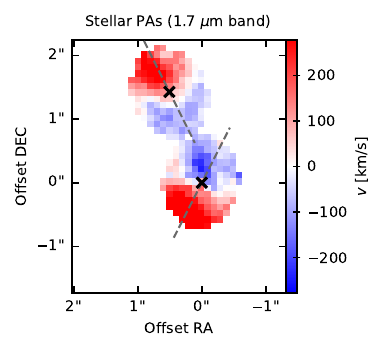}\\
    \caption{Analysis of the stellar dynamics, showing the presence of two rotating structures. The left and middle panels show respectively the stellar velocity and velocity dispersion maps obtained from the 1.0 $\mu$m and 1.7 $\mu$m bands using pPXF (we note that the colour bars have different limits). The points N, S, C, and OE indicated on the maps are the same as Fig.~\ref{fig:rgb}. The other two positions are centred at R.A.=16$^h$52$^m$58$^s$.829, Dec.=02$^{\circ}$24$'$4$''$.369 (ONW) and 
    R.A.=16$^h$52$^m$58$^s$.863, Dec.=02$^{\circ}$24$'$2$''$.331 (OSW). Examples of pPXF spectral fits are presented in \cref{fig:ppxf-results}. The right panel shows the major axis PA (gray dashed lines) for the two stellar structures derived using the \texttt{kinemetry} tool \citep{krajnovic_kinemetry_2006}. The two AGN, marked with crosses, are used as kinematic centres. The northern PA is $26.5\pm3.2^\circ$, and the southern PA is $153.0\pm0.5^\circ$. Angles are measured counter-clockwise from the north.}
    \label{fig:stellar_kin}
\end{figure*}

We fit a stellar continuum to the data using the penalized pixel-fitting routine (pPXF)  \citep{Cappellari+Emsellem2004, cappellari17}. We use the E-MILES stellar population synthesis templates \citep{vazdekis_uv-extended_2016}, which cover the wavelength range $0.168-5\,\mu$m at a spectral resolution of $2.51-23.57$ Å. Since the shape of the stellar continuum varies between the bands, we adopt different orders of Legendre polynomials to correct the shape during the fit. We use a fifth-order additive and a second-order multiplicative Legendre polynomial for the 1.0 $\mu$m band. We also use a fifth-order additive polynomial for the 1.7 $\mu$m band, with a third-order multiplicative polynomial to account for the shape variations. The 2.9 $\mu$m band is difficult, as there are few stellar features for pPXF to work with. We attempt a fit with a sixth-order additive and a third-order multiplicative polynomial, but the comparison with the results obtained in the other two bands indicates that the stellar fit results are unreliable at these longer wavelengths. 

In \cref{fig:stellar_kin}, we present the stellar velocity and velocity dispersion maps obtained using pPXF for the 1.0 $\mu$m and 1.7 $\mu$m bands. In the maps, we marked the positions of six representative spaxels across the FoV (including the two nuclei) characterised by different levels of S/N, for which we show examples of pPXF spectral fits in \cref{fig:ppxf-results}. The resulting  stellar moment maps are consistent with previous studies \citep[e.g.,][]{ceci_jwstnirspec_2024,Engel+10,Ilha+16,Kollatschny+20}, especially those obtained from the 1.7 $\mu$m band data which we deem to be more reliable thanks to the availability of numerous stellar features. 
The rightmost panel of \cref{fig:stellar_kin} shows the result of a fit to the 1.7 $\mu$m band stellar maps with the \texttt{kinemetry} tool \citep{krajnovic_kinemetry_2006}, obtained by selecting the VLBI positions of the two AGN as kinematic centres. Such kinemetry fit is shown only for better visualising the structure and orientation of the two stellar rotating structures, and it will not be used in the analysis that follows. \cite{Engel+10} already performed a detailed analysis of the stellar dynamics, exploring different choices of kinematic centres and using data with a higher spectral resolution.

\cref{fig:stellar_kin} confirms the presence of velocity gradients suggestive of two distinct rotating structures centred at the N and S nuclei. According to \cite{Engel+10}, their properties are consistent with the (pseudo-)bulges of the progenitor merging disc galaxies. As already pointed out by previous studies, the stellar velocity dispersion does not display any clear peak at the expected dynamic centres. However, especially in the 1.7~$\mu$m band map, we can identify regions of high $\sigma_v$
between the S~AGN and position C, with peak values of $\sigma_v>350$~km~s$^{-1}$.
Although the limited spectral resolution of the JWST data demands some caution, a similar enhancement of $\sigma_v$ at the northern side of the southern stellar pseudo-bulge is visible in the stellar moment maps reported by \cite{Engel+10}. These were obtained with the SINFONI instrument at the Very Large Telescope (VLT) equipped with adaptive optics and delivering an angular resolution of $0.10''\times0.16''$. These authors demonstrated that the $\sigma_v>330~$~km~s$^{-1}$ values can be the consequence of a single-kinematic component fit performed on data where absorption features from two stellar populations at different relative line-of-sight velocities are blended. The two likely populations are the southern progenitor's stars and younger stars formed during the collision. Their estimates indicate that about 6\% of the extinction-corrected luminosity in this central region of NGC~6240 may trace a younger stellar population formed outside of the progenitor's pseudo-bulge.

\subsection{Spectral line fitting}\label{sec:results_spectralfit}

\begin{table}[tbp]
\caption{Fitted emission lines}\label{tab:fitted-em-lines}
\centering
 \begin{tabular}{lcc}
    \hline\hline
    Transition & $\lambda_{vac}$ [$\mu$m] & Fitting region [$\lambda_l$,$\lambda_u$]  [$\mu$m] \\
    \hline
    $\rm{Pa\alpha}$ & 1.8756 & $[1.8578,1.8880]$  \\
    $\rm{Pa\beta}$ & 1.2822 & $[1.2690,1.2940]$ \\
    $\rm{[Fe\,II]\lambda1.257\mu m}$ & 1.2567 & $[1.2500,1.2640]$ \\
    $\rm{[Fe\,II]\lambda1.644\mu m}$ & 1.6435 & $[1.6300,1.6590]$ \\
    H$_2$ 1-0 S(1) & 2.1218 & $[2.1100,2.1350]$ \\
    H$_2$ 0-0 S(8) & 5.0529 & $[5.0200,5.0900]$ \\
    \hline
   \end{tabular} 
\end{table}

\begin{table*}[tbp]
\begin{threeparttable}
    \centering
    \caption{Initial parameters and parameter bounds used for the three Gaussian functions during the spectral line fitting.}
    \small
    \begin{tabularx}{\textwidth}{lX | XcX | XcX | XcX | XcX | XcX | XcX}
        \hline
        \hline
        \multirow{2}{*}{Component} && \multicolumn{9}{c|}{Initial parameter value} & \multicolumn{8}{c}{Parameter bounds} \rule{0pt}{2.1ex}\rule[-0.9ex]{0pt}{0pt}\\
        \cline{3-19}
        &&& $A$ &&& $\mu_0$  &&& $\sigma_{\lambda}$  &&& $A$ &&& $\mu_0$  &&& $\sigma_{\lambda}$  \rule{0pt}{2.1ex}\rule[-0.9ex]{0pt}{0pt}\\
        \hline 
        Rotating comp. &&& 0.3 &&& $\mu_*$ &&& $\sigma_*$ &&& $0.1-1.0$ &&& $\mu_{*,l}-\mu_{*,u}$ &&& $\sigma_{*,l}-\sigma_{*,u}$ \rule{0pt}{2.1ex} \\
        Non-rotating comp. 1 &&& 0.5 &&& $\lambda_0-0.001$ &&& 0.001 &&& $0.0-1.0$ &&& $\lambda_\rm{min}-\lambda_\rm{max}$ &&& $0.0-0.003$ \\
        Non-rotating comp. 2 &&& 0.0 &&& $\lambda_0+0.001$ &&& 0.001 &&& $0.0-1.0$ &&& $\lambda_\rm{min}-\lambda_\rm{max}$ &&& $0.0-0.003$ \\
        \hline
    \end{tabularx}
    \vspace{0.9ex}
    \label{tab:fit-params}

    {\it Notes:} Amplitudes ($A$) are peak normalized; central wavelengths ($\mu$) and line widths ($\sigma_{\lambda}$) have the same units $[\mu m]$. The initial values and bounds for $\mu$ and $\sigma_{\lambda}$ of the rotating component are determined from the results of the \texttt{pPXF} fit on the stellar features in the corresponding spaxel (see Sect.~\ref{sec:results_spectralfit} for a detailed explanation of our fitting approach). $\lambda_0$ is the expected observed central wavelength of the emission lines computed using $z=0.02448$. $\lambda_{\rm{min}}$ and $\lambda_{\rm{max}}$ define the spectral range considered in the fit.
\end{threeparttable}
\end{table*}

In Sect.~\ref{sec:results_stellarkin} we have shown that the stellar light of NGC~6240 within the NIRSpec FoV is dominated by two rotating structures, one around each nucleus. In addition, we know from previous works that NGC~6240 hosts a powerful multi-phase galactic outflow, embedding significant molecular gas. As pointed out by several studies, the massive H$_2$ outflow is probably the reason why, contrary to ionised gas tracers and to the stellar component, all molecular tracers in this source display a surface brightness peak between the two nuclei and do not show clear signs of rotation (see Sect.~\ref{sec:Intro}).

A primary goal of our analysis is to disentangle the gas components involved in the outflow from those that are not. To do so, we cannot rely on standard disc modelling procedures followed by a study of residual emission: such approach is valid for sources where disc-like rotation dominates the gas emission, but this is not NGC~6240's case. Another approach could be a narrow/broad spectral line decomposition such as the one adopted by \cite{ceci_jwstnirspec_2024}. However, we note that this commonly used method relies on the assumption that the bulk of the flux from a given emission line is in a ``narrow'' spectral component tracing a rotating/disc structure, while secondary broad and/or high-velocity components trace an outflow (or other non-rotating structures). This assumption may be valid for less extreme galaxies, but it does not hold for NGC~6240, where the bulk of the line emission, especially from molecular gas tracers, does not trace rotation. It is instead likely that non-virial motions and, in particular, outflows, dominate the gas kinematics in the central $\sim2$~kpc of this galaxy (the NIRSpec FoV). This is supported by the highly blueshifted OH absorption \citep{Veilleux+13} and by
the CO and [CI] analysis presented in \cite{Cicone+18}, which had the advantage of using (sub-)mm data with extremely high spectral resolution and S/N. Further confirming this hypothesis, an inspection of the moment maps divided into ``narrow'' and ``broad'' components presented by \cite{ceci_jwstnirspec_2024} for the [FeII], Pa$\alpha$, and H$_2$ 1-0 S(1) lines shows that: (i) their ``broad'' components maps appear like a lower S/N version of the ``narrow'' components maps, there are no clear kinematic differences between the two, indicating that the outflow has not been properly de-blended; (ii) none of the ``narrow'' component maps resemble the rotating pattern seen in the stars. In fact, the ``narrow'' velocity maps of all three gas tracers display a highly redshifted area between the two AGN, a region where the rotating stellar structures are instead systematically blue-shifted. Not surprisingly, this is a region where the outflow brightness is maximum (see later Section~\ref{sec:Discussion}), clearly proving that the outflow is not just contaminating, but it is even dominating the flux in the so-called ``narrow'' components identified by \cite{ceci_jwstnirspec_2024}.

We thus devised a new fitting approach to disentangle rotating from non-rotating motion that relies on the results of the stellar kinematics fit. Given the gas-rich nature of the progenitor disc galaxies involved in the merger, and the rotation pattern dominating the stellar kinematics, we assume that there must be some gas coupled with the stars in such rotating structures. This assumption is supported by the detection of spatially concentrated (sub-)mm dust emission around both AGN, which implies the presence of $\sim10^8~M_{\odot}$ and $\sim8\times10^8~M_{\odot}$ of circumnuclear molecular gas at $r\lesssim40$~pc \citep{Medling+19}. These scales are well within the stellar rotating structures. CO(2-1) emission has also been detected on the same scales \citep{Medling+19}.
Under this assumption, we tie one spectral component of the gaseous emission lines to the stellar kinematics, forcing it to share the same velocity and velocity dispersion as the stars (on a spaxel-by-spaxel basis) within their $3\sigma$ uncertainty derived from \texttt{pPXF}. Up to two additional emission line components will then account for the gas emission that is decoupled from the stars. Assuming the same $\sigma_v$ of rotation for gas and stars is a limitation of this approach: however, given the low spectral resolution of the NIRSpec data, and the absence of a clear outflow-free region in the FoV that would allow us to isolate the gas involved in the rotation and measure its $\sigma_v$ (without the outflow contamination), this is the best that can be done for this source. Furthermore, assuming that in each spaxel each stellar component has an associated gas component may lead to overestimating the fraction of gas coupled with the stars, and consequently underestimating the fraction of gas not involved in stellar motions. Still, this is a clear improvement from the common practice of using only the high-velocity wings of emission (or absorption) lines to trace outflows or inflows, which leads to severe underestimation of these components.
Despite its limitations, our methods are more physically motivated and more tailored to the complex morphology and kinematics of a major galaxy merger such as NGC~6240. We note that a similar approach has been successfully implemented by \cite{Hagedorn+26} to study the multiphase outflow in IRAS~20100-4156.

Henceforth, we will define the gas component that is kinematically coupled to the stars as ``rotating'', as opposed to ``non rotating'' components. We note that this is an oversimplification, because by construction these components account for any motion with a stellar counterpart in the NIRSpec bands. For example, gas-rich stellar tidal tails and similar merger-related features are included among such ``rotating'' components.
The high dispersion ($\sigma_v>350$~km~s$^{-1}$) region between the two AGN discussed at the end of Sect.~\ref{sec:results_stellarkin} is one of such cases.

For the purpose of rotating/non-rotating components decomposition, we select six bright and representative emission lines from the rich NIR spectrum of NGC~6240 and fit them on a spaxel by spaxel basis,  using the \texttt{lmfit} Python package for non-linear curve fitting \citep{newville_lmfitlmfit-py_2024}.
We choose two hydrogen recombination lines ($\rm{Pa\alpha}$ \& $\rm{Pa\beta}$), two iron lines ($\rm{[Fe\,II]\lambda1.257\mu m}$ \& $\rm{[Fe\,II]\lambda1.644\mu m}$), and two molecular hydrogen lines; one ro-vibrational (H$_2$ 1-0 S(1)) and one pure rotational (H$_2$ 0-0 S(8)). \cref{tab:fitted-em-lines} reports the central wavelengths and the spectral regions considered during the fits.
We perform an $\rm{S/N}$-cut by taking the median of the intensity at each spaxel and only choosing the spaxels with $\rm{S/N>10}$. Then, each spectrum is fitted using a compound model consisting of up to three Gaussian functions and one linear function. As described above, one Gaussian shares similar $v$ and $\sigma_v$ as the stellar absorption features in that given spaxel as obtained from \texttt{pPXF}, with values that are allowed to vary up to within 3 times the 1-sigma error-bars derived from \texttt{pPXF}. The other (one or two) Gaussians are free to trace the remaining spectral emission arising from gas whose dynamics is not consistent with that of the stars. The linear function accounts for any baseline residuals after the stellar continuum subtraction. 

The fit residuals are calculated as $\rm{res.}=w(\rm{data-fit})$, where $w$ are the weights. We set the weights to be the inverse spectral RMS noise level. We set the minimization optimizer to Powell's method, a conjugate direction method, which we find to be better suited for multi-component spectral fitting with several local minima than the default optimizer (the Levenberg-Marquardt method). For each spaxel, we scale the spectrum by the maximum intensity to avoid numerical instabilities while performing each fit. 
\cref{tab:fit-params} reports the initial values and parameter bounds used in our fit. 
The spectral fit results for the selected spaxels marked in \cref{fig:stellar_kin} are shown in Figs~\ref{fig:pa-fit} to \ref{fig:fe-fit}. 
The total line fluxes, integrated over the full NIRSpec FoV, are reported in \cref{tab:line-fluxes}, together with the line fluxes measured separately for the rotating and non-rotating components. 

\begin{table}[tbp]
\centering
\caption{Line fluxes measured from the fit.}
\label{tab:line-fluxes}
\begin{tabular} {lccc}   
\hline
\hline
Line & $F_\rm{line}^\rm{tot.}$ & $F_\rm{line}^\rm{rot.}$  & $F_\rm{line}^\rm{nonrot.}$ \\
    & \multicolumn{3}{c}{[$\rm{10^{-15}erg\,s^{-1}cm^{-2}}$]} \\
\hline
    Pa$\alpha$ & $71.1\pm0.4$ & $25.43\pm0.16$ &  $45.6\pm0.4$  \\ 
    Pa$\beta$ & $30.5\pm2.6$ & $\phantom{0}9.6\pm0.3$  & $20.9\pm2.5$  \\
    H$_2$ 1-0 S(1) & $57.5\pm0.9$ & $19.7\pm0.5$  & $37.8\pm0.7$ \\
    H$_2$ 0-0 S(8) & $16.2\pm0.1$ & $\phantom{0}8.0\pm0.1$ &  $\phantom{0}8.2\pm0.1$ \\
    \text{[Fe II]}$\lambda1.257\mu$m & $55.9\pm1.1$ & $19.4\pm0.3$ & $36.5\pm1.1$ \\
    \text{[Fe II]}$\lambda1.644\mu$m & $55.9\pm0.4$ & $19.81\pm0.19$ & $36.1\pm0.4$ \\
  \hline
\end{tabular}
\begin{flushleft}
{\it Notes:} $F_\rm{line}^\rm{tot.}$ is the total line flux, integrated across the whole FoV; $F_\rm{line}^\rm{rot.}$ is the line flux measured in the rotating components; $F_\rm{line}^\rm{nonrot.}$ is the sum of fluxes of the Gaussian components (one or two) that account for non-rotating motions. 
\end{flushleft}
\end{table}

\subsection{The PAH 3.3 $\mu$m feature}\label{sec:results_PAH}

\begin{figure}[tbp]
    \centering
    \includegraphics[clip=true,trim=0.1cm 0.3cm 0.3cm 0.2cm, width=0.48\columnwidth]{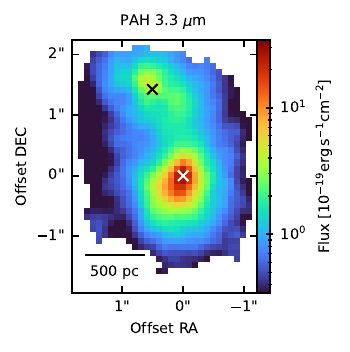}\quad
    \includegraphics[clip=true,trim=0.1cm 0.3cm 0.3cm 0.2cm, width=0.48\columnwidth]{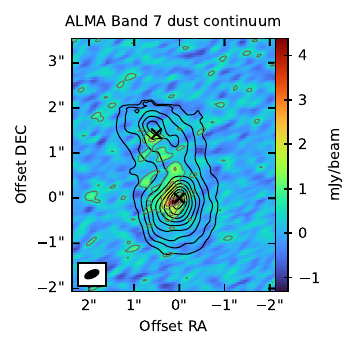}\\
    \caption{{\it Left panel:} Intensity map of the PAH 3.3 $\mu$m band. The AGN positions are marked with crosses. {\it Right panel:} High resolution ALMA map of the 870$\mu$m dust continuum, with overlaid PAH 3.3$\mu$m contours. }
    \label{fig:pah-int}
\end{figure}

When exposed to UV radiation, the Carbon-rich molecules consisting of tens to hundreds of Carbon atoms denominated polycyclic aromatic hydrocarbons (PAHs), get excited as the energy from UV photons is transferred to their vibrational levels. When cooling radiatively, PAHs produce strong, broad emission features at NIR and MIR wavelengths. The PAH 3.3$\mu$m band, whose emission in NGC~6240 is prominent across the whole NIRSpec FoV (see Fig.~\ref{fig:gather}), is dominated by smaller neutral molecules that are resilient to AGN radiation fields \citep{rigopoulou_polycyclic_2024}.

To produce a PAH 3.3~$\mu$m intensity map, we fit this feature using two Gaussians: one for the PAH feature itself and one to account for the H$_2$ 1-0 O(5) emission line at 3.235 $\mu$m that is partially blended with the PAH feature. The spectral fit results are reported in the Appendix (\cref{fig:pah-spaxels-fits}). In the left panel of \cref{fig:pah-int}, we show the map of the total integrated intensity of the PAH~$3.3\mu$m feature, which peaks at the S~AGN and extends in a peanut shape with a secondary peak near the N~AGN. The PAH morphology differs from that of the gaseous tracers investigated in this paper, in particular from that of H$_2$ line emission. The UV-pumped IR-fluorescence of PAH molecules potentially follows more closely the distribution of the (sub-)mm dust continuum, shown in the right panel of \cref{fig:pah-int}. 

\subsection{Total moment maps of gas tracers}\label{sec:results_momentmaps}
\begin{figure*}[tbp]
    \centering
    \includegraphics[clip=true,trim=0.1cm 0.3cm 0.3cm 0.2cm, width=\textwidth]{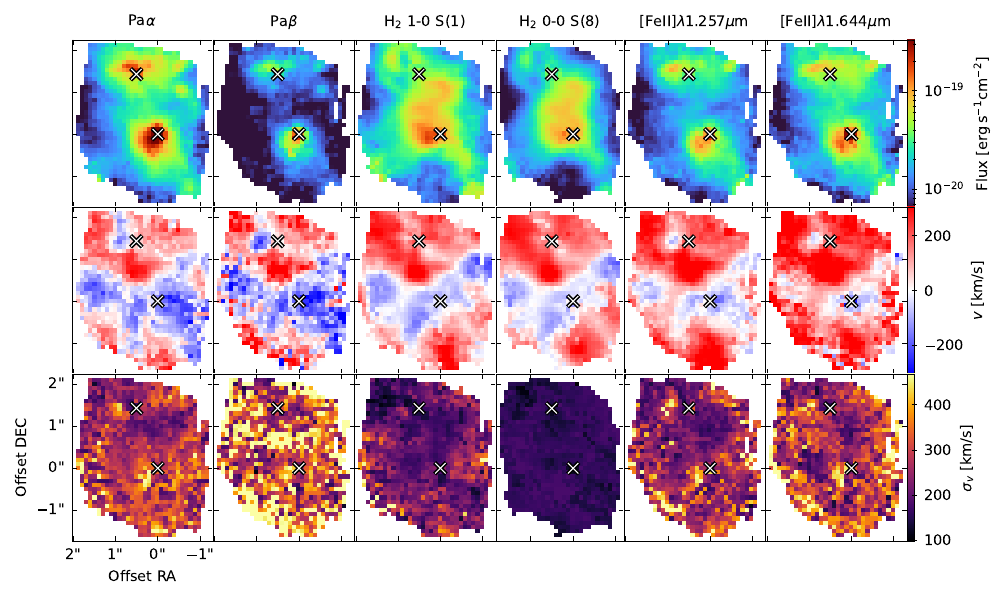}
    \caption{Moment maps for six selected emission lines; two hydrogen recombination lines Pa$\alpha$ and Pa$\beta$, two strong molecular hydrogen lines H$_2$ 1-0 S(1) and H$_2$ 0-0 S(8), and two ionized iron lines [Fe~II]$\lambda1.644\mu$m and [Fe~II]$\lambda1.257\mu$m. The zeroth moment is shown in the upper panel, the first in the middle, and the second in the lower panel. AGN positions are marked with crosses.}
    \label{fig:allmoms}
\end{figure*}

Before analysing separately the rotating and non-rotating gaseous components decoupled by our spectral fit procedure, we first present in \cref{fig:allmoms} the {\it total} 0th moment (intensity), the 1st moment (intensity-weighted velocity), and the 2nd (intensity-weighted velocity dispersion) maps of the six selected emission lines. These maps have been computed considering the total line flux from these transitions, hence, rotational and non-rotational motions are blended in them.

It is evident from the 1st moment maps in \cref{fig:allmoms} that the gas kinematics do not resemble that of the stars. This implies that, in NGC~6240, the bulk of the emission line luminosity, for both molecular and ionised gas tracers, probes gas motions that are decoupled from the stars. The Paschen lines show localized peaks near the nuclei, with a main peak in correspondence with the S~AGN and a secondary peak slightly north-east of the N~AGN. The northern intensity peak is coincident with a prominent blue-shifted component visible in both Pa$\alpha$ and Pa$\beta$, and it is near a region of enhanced velocity dispersion in the Pa$\alpha$ 2nd-moment map (the Pa$\beta$ velocity dispersion map is too noisy to draw this conclusion). A possible higher-$\sigma_v$ counterpart is also seen on the western side of the nucleus, in the Pa$\alpha$ map. Such blue-shifted and high-$\sigma_v$ features could be the signature of a bi-conical ionised wind launched from the northern nucleus in a direction aligned with the minor axis of stellar rotation, which will be discussed in Sect.~\ref{sec:Discussion}.
Overall, the Pa$\alpha$ and Pa$\beta$ velocity maps are similar. The velocity dispersion of Pa$\beta$ is much higher than for any of the other lines, which can be due to its lower S/N.

In the 0th moment maps of the $\rm{H_2}$ lines, the surface brightness is enhanced across an extended, approximately rectangular ($1.0''\times 1.5''$, i.e. 760~pc $\times$ 510~pc) region between the two AGN, with a peak near the S~AGN, a bit shifted to the east. The morphology is similar for both H$_2$ lines. Such concentration in the internuclear region is a distinctive property of NGC~6240 and it is shared by all H$_2$ tracers investigated so far (see Sect.~\ref{sec:Intro}).
The velocity maps of the ro-vibrational and rotational H$_2$ transitions are almost identical, with the N~AGN embedded in redshifted gas with $v_\rm{H_2}>150$ km/s, and the S~AGN situated in a blue-shifted diagonal band in the south-east/north-west direction with $v_\rm{H_2}\sim-50$ km/s. 

The intensity maps of the two [FeII] lines show similar features as the two Paschen lines, with peaks near the two nuclei. However, the [FeII] velocity maps are dominated by redshifted gas, with a small region of blue-shifted gas north-east of the N~AGN, similar to what is seen in Pa$\alpha$ and Pa$\beta$. Akin to the other lines, the [FeII] gas is blue-shifted in the area surrounding the S~AGN. The velocity dispersion map of both [Fe~II] tracers shows similarities with the Pa$\alpha$ moment-2 map. 

\subsection{Moment maps of non-rotating gas components}\label{sec:results_nonrot} 

\begin{figure*}[tbp]
    \centering
    \includegraphics[clip=true,trim=0.1cm 0.3cm 0.3cm 0.2cm, width=\textwidth]{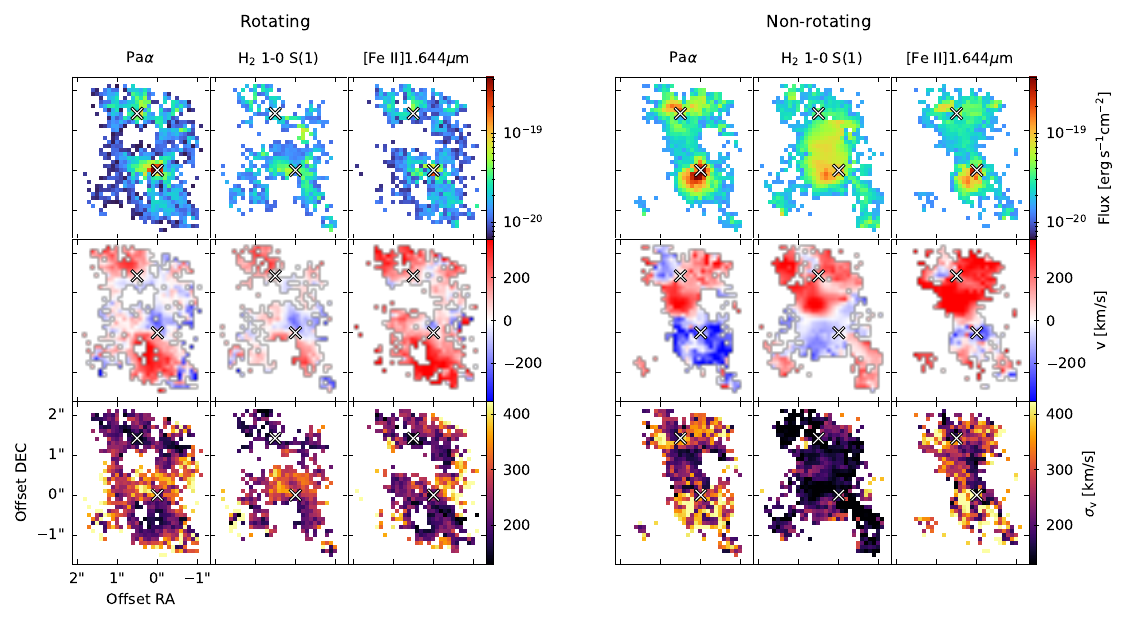}
    \caption{Moment maps of rotating ({\it left}) and non-rotating ({\it right}) components of the $\rm{Pa\alpha}$, H$_2$ 1-0 S(1), and $\rm{[Fe\,II]\lambda1.644\mu}$m emission lines. Only spaxels with S/N$>2$ are plotted. The crosses indicate the AGN positions.}
    \label{fig:disc-and-nondisc-moments}
\end{figure*}

Our line fitting procedure, described in Sect.~\ref{sec:results_spectralfit}, allowed us to de-blend the non-rotating from the rotating components of the gas emission. We remind that the latter  include by construction any component with a stellar counterpart in the NIRSpec data, such as merger-related features. The resulting moment maps are reported in \cref{fig:disc-and-nondisc-moments} for the
$\rm{Pa\alpha}$, H$_2$ 1-0 S(1), and $\rm{[Fe\,II]\lambda1.644\mu m}$ emission lines, i.e. the three transitions with the highest S/N and so the most reliable spectral fit results.
As expected, the rotating components, presented in the left panels of \cref{fig:disc-and-nondisc-moments}, follow closely the stellar kinematics. The moment maps of the non-rotating components, presented in the right panels, are very similar to the total moment maps (Fig.~\ref{fig:allmoms}). This indicates that rotation has a minor contribution to the NIR line emission in these central $\sim2$ kpc of NGC~6240. We find that, within the NIRSpec FoV, a rotating component accounts for $35.8\pm0.4$\% of the Pa$\alpha$ flux, $34.3\pm1.1$\% of the H$_2$ 1-0 S(1) flux, and $35.5\pm0.5$\% of the $\rm{[Fe\,II]\lambda1.644\mu}$m flux (see \cref{tab:line-fluxes}).

The non-rotating component of $\rm{Pa\alpha}$ near the N~AGN has a shape that is elongated approximately perpendicular to the major-axis of stellar rotation. Both the blue-shifted eastern side and the red-shifted western sides display a high velocity dispersion. This could already be distinguished (at a lower S/N) in the total moment maps in Fig.~\ref{fig:allmoms}. We interpret this feature as a bi-conical wind expanding along the minor axis of rotation. We detect a similar feature in the non-rotating $\rm{[Fe\,II]\lambda1.644\mu m}$ moment 2 map, with a faint counterpart also in the non-rotating H$_2$ 1-0 S(1) moment 2 map. 
Because of its characteristic geometry, its high-velocity dispersion of 400~km~s$^{-1}$, and because it is clearly decoupled from the stars, we suggest that this structure traces an AGN-driven wind entraining ionised gas and warm molecular gas, extending out to several 100s~pc from the N~AGN. 

In the southern half of the FoV, around and south of the S~AGN, the moment maps of the non-rotating components of all three lines show an extended area of enhanced velocity dispersion, characterised by blue-shifted velocities. This feature does not show a clear elongation or orientation with respect to stellar rotation, so it is hard to reconcile it with a classical bi-conical wind similar to that identified around the N~AGN. We discuss it further in Sect.~\ref{sec:Discussion}.

\subsection{H$_2$/PAH ratio map}\label{sec:results_H2PAH}
\begin{figure}[tbp]
    \centering
    \includegraphics[clip=true,trim=0.1cm 0.3cm 0.2cm 0.2cm, width=0.48\columnwidth]{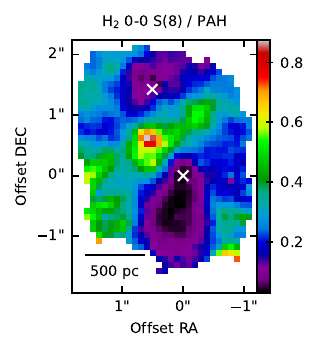}\quad
    \includegraphics[clip=true,trim=0.1cm 0.3cm 0.2cm 0.2cm, width=0.48\columnwidth]{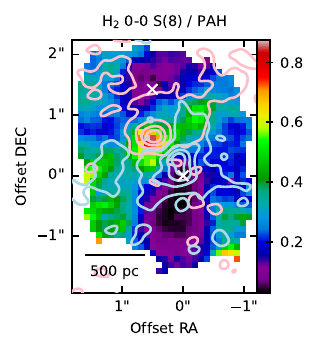}\\
    \caption{Map of the intensity ratio between the rotational H$_2$ 0-0 S(8) line and the PAH 3.3 $\mu$m emission band. The contours in the right panel are the [CI](1-0) red- and blue-wing outflow emission from \cite{Cicone+18}, corresponding to $[3\sigma, 6\sigma, 15\sigma, 21\sigma]$ with $\sigma=1.1$ mJy/beam for the red wing, and 1.0 mJy/beam for the blue wing. White crosses mark the AGN positions.}
    \label{fig:H2_PAH_ratiomaps}
\end{figure}    

Excess H$_2$/PAH emission, by a factor of 10-100 compared to the disc, has been found in the galactic outflow of M82 and explained with shock excitation 
\citep{Beirao+15}.
In \cref{fig:H2_PAH_ratiomaps} we present
the H$_2$ 0-0 S(8)/PAH 3.3$\mu$m intensity ratio map of NGC~6240. We obtain similar maps if we compute the H$_2$/PAH ratio using a ro-vibrational H$_2$ line instead of H$_2$ 0-0 S(8). Both AGN are located in regions where the H$_2$/PAH ratio is low. This is due to the conjunction of PAH 3.3$\mu$m being stronger at the positions of the AGN (Fig.~\ref{fig:pah-int}), and H$_2$ emission being brighter between them. We can distinguish two low H$_2$/PAH ratio ($<0.2$) areas centred on the two AGN, both extending away from the nuclear region, i.e. south of the S~AGN and north/north-west of the N~AGN. 
Interestingly, the AGN wind that we identified in the northern nucleus coincides with one of such low H$_2$/PAH ratio zones. 
Between the AGN, approximately $1''$ south of the N~AGN and slightly offset to the east, there is a compact region where the H$_2$/PAH ratio peaks ($>0.75$).  
The right panel of \cref{fig:H2_PAH_ratiomaps} displays the same H$_2$ 0-0 S(8)/PAH intensity ratio map where we overlaid as contours the red- and blue-shifted outflow emission from cold molecular gas as traced by the ALMA [CI](1-0) data from \cite{Cicone+18}. The [CI](1-0) outflow, especially its redshifted side and its more extended blue-shifted eastern component, follows quite closely the north-west to south-east stripe of elevated H$_2$-to-PAH ratio revealed by the JWST data. Remarkably, the [CI](1-0) red-wing peaks exactly on top of the H$_2$/PAH ratio peak mentioned above.

Taken together, these results suggest that: (i) the bulk of the H$_2$ emission in the internuclear region does not arise from photon-dominated regions (PDRs), because in this case we should detect a peak also in PAH emission due to the excitation of these molecules by UV photons; (ii) the excess H$_2$ emission (responsible for the overall high H$_2$/PAH ratio) originates in the outflow, and the outflow either does not contain PAH grains or, alternatively, it is shielded from UV radiation hence reducing the PAH heating efficiency; (iii) the H$_2$ excitation in the outflow is not due to UV-pumping, but it is thermal (see also \cite{ceci_jwstnirspec_2024}), likely produced by shocks. A shock origin for the NIR H$_2$ emission in NGC~6240 was pointed out by previous studies based on several lines of evidence, such as the high H$_2$/Br$\gamma$ ratios (3-48, \citealt{muller-sanchez_two_2018}), the H$_2$ morphology, and high-velocity H$_2$ wings \citep{Ohyama+00, vanderWerf+93}. \cite{vanderWerf+93} proposed that the shocked H$_2$ emission is due to the ISM collision in the merger, with a contribution from the outflow for the highest velocity H$_2$ gas. As further argued in Sect.~\ref{sec:Discussion}, based on our results, we strongly favour a scenario where most of the bright H$_2$ emission in the nuclear region of NGC~6240 is due to cooling of shocked outflowing gas.

\section{Discussion}\label{sec:Discussion}

\subsection{Widespread outflows dominating the gas kinematics}\label{sec:disc_outflows}
\begin{figure*}[t]
    \centering
\includegraphics[clip=true,trim=0.1cm 0.3cm 0.3cm 0.2cm, width=0.29\textwidth]{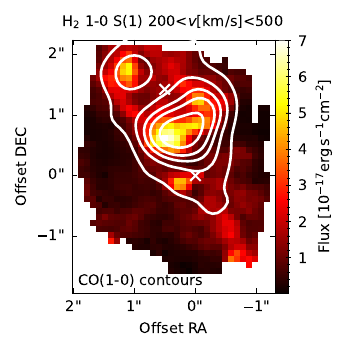}\quad
\includegraphics[clip=true,trim=0.1cm 0.3cm 0.3cm 0.2cm, width=0.29\textwidth]{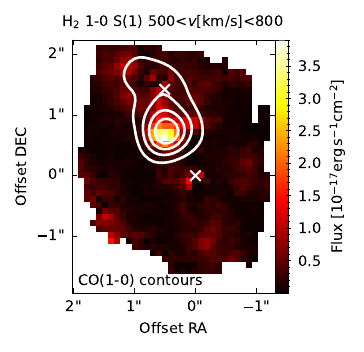}\\
\includegraphics[clip=true,trim=0.1cm 0.3cm 0.3cm 0.2cm, width=0.33\textwidth]{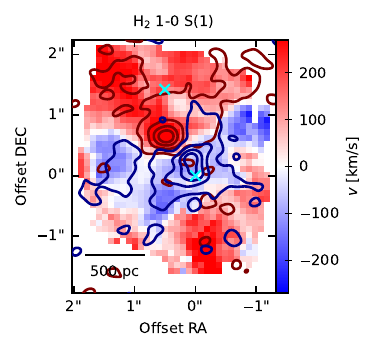}\quad
\includegraphics[clip=true,trim=0.1cm 0.2cm 0.2cm 0.2cm, width=0.45\textwidth]{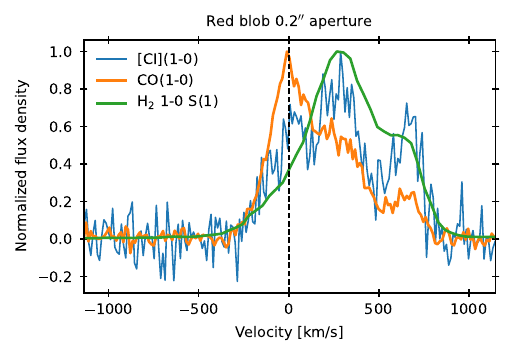}\\
\caption{\textit{Top left:} H$_2$ 1-0 S(1) emission integrated within $v\in[200,500]$ km/s. Overlaid contours show the CO(1-0) outflow emission integrated within the same velocity range, corresponding to $[60\sigma,100\sigma,140\sigma,180\sigma]$ with $1\sigma=0.05\,\rm{Jy\,km\,s^{-1}beam^{-1}}$ . \textit{Top right:} H$_2$ 1-0 S(1) emission integrated within $v\in[500,800]$ km/s. Overlaid contours show the CO(1-0) outflow emission within the same velocity range, corresponding to [$6\sigma, 12\sigma,18\sigma,24\sigma$] with $1\sigma=0.07\,\rm{Jy\,km\,s^{-1}beam^{-1}}$. 
{\it Bottom left:} Total H$_2$ 1-0 S(1) velocity map (i.e. same as shown in Fig.~\ref{fig:allmoms}), with overlaid contours corresponding to the [CI](1-0) red- and blue-wing outflow emission. The crosses mark the AGN positions. \textit{Bottom right:} Peak-normalized [CI](1-0), CO(1-0), and H$_2$ 1-0 S(1) spectra extracted from an aperture width diameter $\sim0.2''$ centred on the peak of highly redshifted molecular gas emission (``red blob'') seen in the two maps in the top panels, at coordinates RA=16:52:58.9219, Dec=02:24:04.00355.
}
    \label{fig:h2-vel-co2-0}
\end{figure*}

\begin{figure}
    \centering
    \includegraphics[clip=true,trim=0.2cm 0.3cm 0.2cm 0.2cm, width=0.4\textwidth]{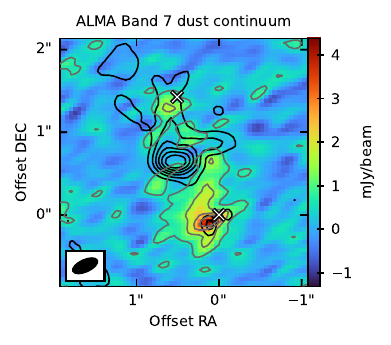}\\
    \caption{ALMA Band 7 (870 $\mu$m) continuum map with overlaid redshifted H$_2$ 1-0 S(1) emission integrated within $v\in[500,800]$~km/s (same as top-right map in Fig~\ref{fig:h2-vel-co2-0}). The gray contours showing the band 7 continuum correspond to $[2\sigma, 5\sigma, 8\sigma, 11\sigma]$ with $1\sigma=0.298$ mJy/beam. The synthesized beam of the ALMA data is shown in the bottom left corner. The crosses indicate the AGN locations.}
    \label{fig:Band7continuum_withredshiftedOF}
\end{figure}

\cite{Cicone+18} suggested that the molecular outflow originates from between the AGN of NGC~6240, extending toward the east and west directions, and connecting with the larger-scale molecular outflowing material extending by up to 10~kpc across the source, especially in the eastern direction, as observed in CO(1-0), CO(2-1), and [CI](1-0) \citep{Feruglio+13a, Cicone+18}. 
In this work, we used JWST NIRSpec observations of the central region of NGC~6240 to understand to what extent the gas follows the stars in gravitational motions tied to the two progenitors' stellar pseudo-bulges. We relied on a spectral line fitting decomposition (see Sect.~\ref{sec:results_spectralfit}) to identify gas components that are decoupled from the stellar kinematics and so involved in non-rotational motions that lack a stellar counterpart, such as outflows or inflows. This required assuming that a fraction of the emission from gas tracers arises from a component that follows the stellar kinematics. However, this assumption may not hold for all regions of the FoV, and so our estimates of the fraction of gas coupled with the stars should be considered as upper limits.
Our results, despite being somewhat conservative as explained above, revealed that at least $\sim65$~\% (see \cref{tab:line-fluxes}) of NIR emission lines in the central 2~kpc of NGC~6240 trace gas components that are significantly decoupled from the stars. Such severe decoupling between gas and stars could already be inferred by the striking similarity between the total moment maps of gaseous tracers (\cref{fig:allmoms}) and the moment maps of the non-rotating components in the right panel of \cref{fig:disc-and-nondisc-moments}. In the latter, we identified two regions of enhanced velocity dispersion: (i) one around the N~AGN, with an elongated structure extending along the minor axis of stellar rotation that is more pronounced in ionised gas tracers (Pa$\alpha$ and [FeII]) than in the H$_2$ 1-0 S(1) line; (ii) an extended region of higher $\sigma_v$ around and south of the S~AGN, spreading across half of the FoV of the NIRSpec data. Our hypothesis is that the first feature (i) traces a bi-conical wind from the N~AGN, and the second one (ii) traces a powerful and massive outflow, likely originating from the more luminous S~AGN. This outflow coincides with that resolved in [CI](1-0) by \cite{Cicone+18} and in CO(2-1) by \cite{treister_molecular_2020}. 
In the following, we discuss these two outflows separately.

\subsubsection{A bi-conical AGN wind from the northern nucleus}\label{sec:northern-outflow}

The total velocity maps of gaseous tracers in \cref{fig:allmoms} display a blue-shifted region east of the N~AGN that is not consistent with the stellar kinematics, which is especially noticeable in ionised gas tracers, and marginally detected in the H$_2$ lines. The corresponding 2nd moment maps show increased velocity dispersion at the same position. The bi-conical morphology of this feature and its alignment along the minor axis of stellar rotation become evident when isolating the non-rotating gas components in the right panels of \cref{fig:disc-and-nondisc-moments}. 
This bi-conical wind mainly involves ionised gas and extends up to $r\sim500$~pc on both sides of the nucleus. The co-spatiality with coronal line emission ([Si VI], [Si VII], [Mg VIII]) and the partial overlap with the AGN ionisation cone identified by \cite{ceci_jwstnirspec_2024}, strongly point to a typical AGN wind. We will refer to this as an AGN wind to distinguish it from the other outflow discussed in Sect.~\ref{sec:southern-outflow}. Interestingly, the N~AGN is, contrary from the southern one, significantly over-massive compared to the local $M_{BH}-\sigma_*$ relation, by one order of magnitude \citep{Medling+19}.

Figure~\ref{fig:H2_PAH_ratiomaps} shows that, in this region,
there is no excess H$_2$ emission with respect to PAH~3.3~$\mu$m. Therefore, either the H$_2$ entrainment in the wind is low, or the wind hosts significant PAH~3.3$\mu$m emission, for example thanks to its high UV irradiation. 
Both statements can be true, conspiring towards a low H$_2$/PAH ratio in this northern AGN wind. To our knowledge, there have not been prior official reports of a molecular wind from the N~AGN, but only of a ionised wind \citep{muller-sanchez_two_2018}. However, the signature of such wind is marginally detected also in the H$_2$ 1-0 S(1) line. Furthermore, the highest resolution ($0.03''\sim15$~pc) CO(2-1) velocity map available, obtained by \cite{treister_molecular_2020}, shows a bi-conical structure of redshifted gas ($v\sim300-400$~km~s$^{-1}$), extending by $r\sim500-750$~pc on each side of the N AGN, and with an orientation consistent with that of the AGN wind discussed here. This feature is distinct from the highly red-shifted H$_2$ gas ``blob'' where the $H_2$/PAH ratio peaks (see Sect.~\ref{sec:results_H2PAH}) and cannot be identified in the ALMA [CI](1-0) outflow maps from \cite{Cicone+18}, probably due to angular resolution limitations. Hence, we conclude that some molecular gas is embedded in this northern wind, but its small spatial extent makes this component negligible compared to the bright H$_2$ emission from the other regions of the FoV. Moreover, the coincidence with the AGN ionisation cone and its orientation hint at a bipolar wind expanding freely along the minor axis of rotation, hence probably not impacting nor shocking much of the surrounding ISM.

\subsubsection{A powerful outflow from the southern nucleus}\label{sec:southern-outflow}

Spatially-extended, non-rotating components decoupled from the stellar kinematics dominate the surface brightness of all NIR gas tracers investigated here (\cref{fig:disc-and-nondisc-moments} and \cref{tab:line-fluxes}). In Pa$\alpha$ and [FeII] emission, they are brightest around the S AGN, with a secondary peak near the N AGN, in correspondence of the AGN wind discussed previously. In the H$_2$ 1-0 S(1) line, non-rotating components are responsible for the characteristic concentration of H$_2$ gas in the internuclear region, with a peak close to the S~AGN.
This central zone between the two AGN: (i) is characterised by a significant H$_2$/PAH enhancement (\cref{fig:H2_PAH_ratiomaps}), which has been linked to galactic outflows that embed shocked gas \citep{Beirao+15}, and (ii) overlaps with the region where the cold molecular outflow originates. Based on this evidence, in the following we propose an interpretation where most of the NIR H$_2$ emission in this inter-nuclear region of NGC~6240 arises from the same massive molecular outflow extensively studied in FIR and (sub-)mm tracers \citep{Feruglio+13a, Feruglio+13b, Veilleux+13, Cicone+18, Saito+18, treister_molecular_2020}. 

 To further explore the connection between cold and warm molecular gas components in this region, we show in the top panels of \cref{fig:h2-vel-co2-0} the H$_2$ 1-0 S(1) channel maps, produced at redshifted velocities, with overlaid CO(1-0) contours probing the same velocity ranges. We remind that the emission from cold molecular gas tracers such as CO and [CI] at these velocities in NGC~6240 has been robustly ascribed to a massive molecular outflow (see Introduction). These maps demonstrate that there is a perfect spatial and kinematical correspondence between the warm and cold H$_2$ gas emissions. The bottom left panel of \cref{fig:h2-vel-co2-0} displays the velocity map of H$_2$, with superimposed contours corresponding to the blue- and red-shifted wings of [CI](1-0) at $v>200$~km~s$^{-1}$ ($>250$~km~s$^{-1}$ for the redshifted gas) which trace the known outflow. There is a striking correspondence between the two tracers, despite the underlying H$_2$ velocity map probes all of the H$_2$-emitting gas reservoir (not just the non-rotating component) while the ALMA [CI] contours show only the high velocity outflow wings. This strengthens our hypothesis that the whole H$_2$ NIR emission is dominated by the same outflow as seen in cold (sub-)mm tracers.

Determining the origin of such massive molecular outflow has proven to be challenging. It may have been generated by the S~AGN, given: (i) the detection of a separate (mostly ionised) wind from the N~AGN (discussed in Section~\ref{sec:northern-outflow}), (ii) the proximity with the S nucleus, and (iii) the overall disturbed kinematics and high-velocity values that affect gas in an extended area surrounding the S~AGN. However, it remains difficult to explain why the brightest part of this outflow is observed about $\sim1''$ (500~pc in projected distance) north-east of the S~AGN, where there is a highly redshifted (maximum velocity of 800~km~s$^{-1}$) ``blob'' of both cold and warm H$_2$ emission, corresponding to the peak of H$_2$/PAH ratio. The spectra of CO(1-0), [CI](1-0), and H$_2$ 1-0 S(1) extracted at the position of this ``red blob'' are shown in the bottom-right panel of \cref{fig:h2-vel-co2-0}, normalized by their respective peaks. All three spectra are very broad and encompass velocities from $-250$~km~s$^{-1}$ to 900~km~s$^{-1}$. The spectra of the cold gas tracers, i.e. CO(1-0) and [CI](1-0), display three peaks: at $v\sim0$~km~s$^{-1}$, $v\sim300$~km~s$^{-1}$, and $v\sim700$~km~s$^{-1}$, while the H$_2$ line only shows two distinct peaks at $v\sim300$~km~s$^{-1}$, and $v\sim700$~km~s$^{-1}$. The latter may be a consequence of the lower spectral resolution of NIRSpec compared to ALMA. Nonetheless, the comparison between the three spectra shows clearly that the redshifted gas at $v>300$~km~s$^{-1}$ contributes to most of the H$_2$ total flux at this position, and it is also significantly bright in [CI](1-0). 

Although the redshifted ``blob'' appears compact in size at the JWST angular resolution, it consists of multiple clumps, spatially resolved by the high resolution ALMA observations presented by \cite{treister_molecular_2020}. The spectra in \cref{fig:h2-vel-co2-0} show that this blob is spectrally resolved into at least three spectral components. Moreover, Fig.~\ref{fig:Band7continuum_withredshiftedOF}, where the H$_2$ emission is displayed as contours on top of the ALMA Band~7 continuum map, shows that there is no continuum point source associated with the redshifted blob. Therefore, it must be a gas-only feature: because of its spatial and spectral properties and, in particular, the high velocities involved, we hypothesise this blob to be connected to the southern outflow discussed here. The lack of overlap between the high-velocity H$_2$ emission and the (brightest regions of the) (sub-)mm dust continuum  in Fig.~\ref{fig:Band7continuum_withredshiftedOF} is also consistent with theoretical predictions that galactic outflows propagate through paths of least resistance \citep[e.g.,][]{Costa+15}.

In the 0th-moment map of the non-rotating H$_2$ components (\cref{fig:disc-and-nondisc-moments}), we detect south of the S AGN two molecular filaments extending to the south-west and south-east directions. The strongest emission in the south-west coincides with the base of an H$\alpha$-filament, suggested to be due to shock ionisation from stellar winds by \cite{muller-sanchez_two_2018}. These H$_2$ filaments are characterised by a line-of-sight velocity close to zero or slightly redshifted in both H$_2$ and [Fe~II], while the Pa$\alpha$ gas is blue-shifted at these positions. 
Between the two filamentary H$_2$ structures, south of the S AGN, there is an extended zone of low H$_2$/PAH ratio ($\lesssim0.2$, see \cref{fig:H2_PAH_ratiomaps}) that extends to the southern edge of the NIRSpec FoV. \cite{ceci_jwstnirspec_2024}, based on the same NIRSpec data, identified here two separate features: a conical, blue-shifted ($v\sim-800$ km/s) molecular outflow, extending south-east of the S AGN, and, a bit offset from it, an AGN ionisation cone extending southward from the AGN. In our interpretation, what these authors isolate as a blue-shifted (single-cone) H$_2$ outflow would be part of the larger-scale southern/central massive molecular outflow discussed in the previous paragraphs. Indeed, we do not find any indication supporting the hypothesis implicitly made by \cite{ceci_jwstnirspec_2024} that this southern blue-shifted component should be regarded as a separate outflow: in our analysis, as well as in the ALMA [CI](1-0) data (bottom-left panel of \cref{fig:h2-vel-co2-0}), this feature appears connected with the other high-velocity blue-shifted and redshifted molecular gas components extending also north of the S AGN.

\subsection{About the H$_2$ excitation: outflow vs ISM collision?}\label{sec:discus_H2excitation}

The nature of the bright NIR H$_2$ emission in NGC~6240 and its excitation mechanism have been extensively discussed in the literature, since the discovery that this source has a $H_2$ 1-0 S(1)-to-IR luminosity ratio that is one order of magnitude higher than the other (U)LIRGs from the sample of \cite{Goldader+95}. Using H$_2$ line ratios diagnostic diagrams, several works (e.g. \cite{vanderWerf+93, Ohyama+00} and, most recently, \cite{ceci_jwstnirspec_2024}) supported a thermal origin for the H$_2$ excitation in NGC~6240, i.e. due to collisions of molecules maintaining the lowest rotational levels of H$_2$ in thermal equilibrium. We independently confirmed their result, measuring the H$_2$~2-1~S(1)/H$_2$~1-0~(S1) line ratios for all spaxels of the JWST NIRSpec map and finding that they cluster around values of 0.1-0.15, as expected for thermal emission with rotational temperature around 2000~K.

There is instead less consensus about the mechanism responsible for the collisional excitation of the H$_2$ gas. Several lines of evidence point to shocks as a main heating mechanism. First of all, the high H$_2$/PAH ratio (see Sect.~\ref{sec:results_H2PAH}). Secondly, the fact that the whole ionised ISM of NGC~6240 shows shock-like excitation, in both NIR \citep{ceci_jwstnirspec_2024} and optical diagnostic diagrams \citep{Kollatschny+20}. Thirdly, as pointed out by \cite{Meijerink+13}, the highly excited CO ladder of NGC~6240 can only be explained by a global shock excitation of its molecular ISM. These authors argued that its exceptionally high CO line-to-continuum ratio rules out both PDRs and XDRs, as most of the absorbed UV or X-ray photons would heat the dust as well as the gas. In addition, the non detection of OH$^{+}$ and H$_2$O$^{+}$ lines in the {\it Herschel} data shown by \cite{Meijerink+13} implies low ionisation fractions hence discarding cosmic rays or X-rays heating. We note that the shielding of most of the ISM gas from the AGN radiation despite the dual AGN nature of NGC~6240 may be explained by the Compton thick nature of both its AGN \citep{Fabbiano+20}. 

Once established that shocks are the culprit for the ISM excitation, including that of molecular gas, we need to understand what is the origin of the shocks. Due to its major merger nature, most of the discussion in the past has focused on large-scale shocks generated in the collision of the nuclear ISM of the two merging galaxies \citep{vanderWerf+93}. Yet, already \cite{vanderWerf+93} noted that the observed H$_2$ velocities of $\sim900$~km~s$^{-1}$ are too high to be reconciled with this scenario, where the two nuclei are colliding at about 150~km~s$^{-1}$. Hence, they suggested that the high-v H$_2$ is due to shocked gas entrained in the outflow, while the low-v H$_2$ is due to merger-driven shocks. However, based on our analysis, we propose that most of the H$_2$ 1-0 S(1) emission in the nuclear region of NGC~6240, as much as $66\pm2$\% of its total flux (\cref{tab:line-fluxes}), arises in a massive molecular outflow launched from the southern nucleus (see Sect.~\ref{sec:southern-outflow}). Secondly, in Fig.~\ref{fig:H2_PAH_ratiomaps} and Fig.~\ref{fig:h2-vel-co2-0}, we showed that the H$_2$/PAH peak and the most redshifted portion of the massive outflow, at velocities as high as $v\sim900$~km~s$^{-1}$, are coextensive. 
Furthermore, also in Fig.~\ref{fig:h2-vel-co2-0}, we demonstrated that there is an almost perfect correlation between the global H$_2$ dynamics, dominated by the brightest emission at lower projected velocities $|v|<200$~km~s$^{-1}$, and the high velocity wings of [CI](1-0): this shows that there is no discontinuity between the lower- and the higher velocity components of H$_2$, which we found to consistently trace the molecular outflow. For all these reasons, we favour a scenario where shock excitation in the molecular outflow is responsible for the global NIR H$_2$ emission from NGC~6240. This supports the findings of \cite{MontoyaArroyave+24} that the global high CO excitation of local (U)LIRGs may be due to widespread massive molecular outflows.

\section{Summary and conclusions}
In this work, we have used data from NIRSpec/JWST to study the kinematic properties of the gas in the central $\sim2$~kpc of the dual AGN NGC~6240. The main goal of our analysis is to study the outflows in this source. We devised a new spectral-line fitting approach to de-blend rotating- and non-rotating gas components in the observed line emission, which is more physically motivated and better tailored to NGC~6240 than previous approaches. 
Our analysis also makes use of archival ALMA observations of [CI](1-0), CO(1-0), and cold dust continuum emission. 
Our main findings are:
\begin{itemize}
    \item The NIR emission lines in NGC~6240 are dominated by gas components whose kinematics is considerably decoupled from that of the stellar component. These contribute to as much as $64\%$ of the Pa$\alpha$ and [FeII]1.644$\mu$m fluxes, and to $66\%$ of the H$_2$ 1-0 S(1) line flux. The rotational and ro-vibrational NIR H$_2$ lines show the most deviation from the stars, with peak emission between the two rotating stellar structures. The moment maps of the ionised gas tracers investigated in this work, i.e., the Pa$\alpha$, Pa$\beta$, and [FeII] lines at 1.257 and 1.644$\mu$m, also show considerable deviations from the stellar distribution. However, unlike the molecular lines, these tracers exhibit clear intensity peaks at the two AGN positions. 
    \item The PAH~3.3~$\mu$m feature and the cold dust emission at 870$\mu$m peak at the two nuclei; hence, they present a very different morphology from the H$_2$ gas. This supports the hypothesis that the bulk of H$_2$ does not trace PDRs, because otherwise it should coincide with PAH emission due to the excitation of these molecules by UV photons.
    \item The moment maps of the non-rotating components of the Pa$\alpha$ and [Fe~II]$\lambda1.644\mu$m lines revealed a bi-conical wind in the northern nucleus, dominated by ionised gas. The wind is likely launched from the N~AGN and it expands along the minor axis of stellar rotation. The western wind cone coincides with the AGN ionisation region identified by \cite{ceci_jwstnirspec_2024}. Although such AGN wind entrains some H$_2$ gas, it does not show a H$_2$/PAH enhancement. Hence, such wind is either significantly UV-irradiated, which would enhance PAH$~3.3\mu$m emission, or it entrains and shocks only a negligible amount of H$_2$ gas compared to the global H$_2$ content of the galaxy. The latter may result from an expansion of the wind along an ISM-free path.
    \item In addition to the northern AGN wind, we propose that most of the NIR H$_2$ line emission from non-rotating components is involved in a massive outflow that extends across the bottom half of the NIRSpec FoV, possibly launched from the southern nucleus. This hypothesis is supported by the striking correlation between the warm H$_2$ gas kinematics and that of the cold molecular outflow traced by [CI] and CO ALMA data. Therefore, this kpc-scale outflow is rich in both warm and cold H$_2$ gas, and it is connected to the extended component of the cold molecular outflow detected out to 10~kpc in low-J CO and [CI](1-0) emission. Contrary to the northern AGN wind, this southern massive outflow, and in particular its receding side that reaches projected velocities as high as $v\sim900$~km~s$^{-1}$ in all molecular gas tracers, is spatially coincident with the region characterised by the strongest H$_2$/PAH enhancement. This suggests that the shocks responsible for such high H$_2$/PAH ratios are due to the outflow rather than to the ISM collision.
\end{itemize}
In conclusion, we showed that the bulk of the NIR H$_2$ emission from the central $\sim2$~kpc of NGC~6240 arises from $T\sim2000$~K shocked gas entrained in a massive molecular outflow, which is cooling down and coexisting with a colder phase probed by low-J CO and atomic carbon ([CI](1-0)) emission. 

\begin{acknowledgements}
The authors thank the referee for their comments which have helped improve the manuscript. This paper makes use of the following ALMA data: ADS/JAO.ALMA\#2013.1.00813.S, ADS/JAO.ALMA\#2015.1.00717.S, and ADS/JAO.ALMA\#2015.1.00003.S. ALMA is a partnership of ESO (representing its member states), NSF (USA), and NINS (Japan), together with NRC (Canada), MOST and ASIAA (Taiwan), and KASI (Republic of Korea), in cooperation with the Republic of Chile. The Joint ALMA Observatory is operated by ESO, AUI/NRAO, and NAOJ. CC thanks the ALMA European Regional Center for providing calibrated MS of the older ALMA datasets. This work is based in part on observations made with the NASA/ESA/CSA James Webb Space Telescope. The data were obtained from the Mikulski Archive for Space Telescopes at the Space Telescope Science Institute, which is operated by the Association of Universities for Research in Astronomy, Inc., under NASA contract NAS 5-03127 for JWST. These observations are associated with GTO program 1265. CC acknowledges funding from the European Union's Horizon Europe research and innovation programme under grant agreement No. 101188037 (AtLAST2). Views and opinions expressed are however those of the author(s) only and do not necessarily reflect those of the European Union or European Research Executive Agency. Neither the European Union nor the European Research Executive Agency can be held responsible for them. CC, CV, PS acknowledge financial support from the INAF Bando Ricerca Fondamentale INAF 2022 Large Grant: ``Dual and binary supermassive black holes in the multi-messenger era: from galaxy mergers to gravitational waves''
and from the INAF Bando Ricerca Fondamentale INAF 2024 Large Grant: ``The Quest for dual and binary massive black holes in the gravitational wave era''.
\end{acknowledgements}

\bibliographystyle{aa}
\bibliography{references_ngc6240}

\appendix
\section{Supplementary material}\label{sec:appendix}

\begin{table*}[tbp]
\begin{threeparttable}
    \caption{Identified emission lines}
    \label{tab:identified-lines}
    \small
    \centering
    \begin{tabular}{c  c | c  c  c  c| c  c  c  c}
    \hline\hline
    \multirow{3}{*}{Ionized gas} & \multirow{3}{*}{$\rm{\lambda_{vac}\,[\mu m]}$} & \multicolumn{8}{c}{Warm molecular gas}\rule{0pt}{2.1ex}\rule[-0.9ex]{0pt}{0pt} \\
    \cline{3-10}
    && \multicolumn{4}{c|}{Ro-vibrational transitions} & \multicolumn{4}{c}{Pure rotational transitions} \rule{0pt}{2.1ex}\rule[-0.9ex]{0pt}{0pt} \\
    \cline{3-10}
    && H$_2$ (1-0) & $\rm{\lambda_{vac}\,[\mu m]}$ & H$_2$ (2-1) & $\rm{\lambda_{vac}\,[\mu m]}$ & H$_2$ (0-0) & $\rm{\lambda_{vac}\,[\mu m]}$ & H$_2$ (1-1) & $\rm{\lambda_{vac}\,[\mu m]}$ \rule{0pt}{2.1ex}\rule[-0.9ex]{0pt}{0pt} \\
    \hline
    [SIII] & 0.9533 & S(7) & 1.7480 & S(3) & 2.0735 & S(16) & 3.5476 & S(9) & 4.9541 \rule{0pt}{2.1ex} \\
    $\rm{[CI]}$ & 0.9853 & S(6) & 1.7881 & S(1) & 2.2477 & S(15) & 3.6262 \\
    $\rm{[SII]}$ & 1.0323 & S(5) & 1.8358 & Q(1) & 2.5509 & S(14) & 3.7244 \\
    HeI & 1.0832  & S(4) & 1.8919 & Q(3) & 2.5698 & S(13) & 3.8461 \\ 
    $\rm{[PII]}$ & 1.1886 & S(3) & 1.9576 & O(3) & 2.9741 & S(11) & 4.1811 \\
    $\rm{[Fe~II]}$ & 1.2567 & S(2) & 2.0338 & O(5) & 3.4379 & S(10) & 4.4098 \\
    & 1.3209 & S(1) & 2.1218 & & & S(9) & 4.6946 \\
    & 1.3720 & S(0) & 2.2233 & & & S(8) & 5.0531 \\
    & 1.6436 & Q(3) & 2.4237 &&\\
    $\rm{Pa\beta}$ & 1.2822 & Q(4) & 2.4375 &&\\
    $\rm{Pa\alpha}$ & 1.8756 & Q(5) & 2.4548  &&\\
    $\rm{Br\gamma}$ & 2.1661 & Q(6) & 2.4756 &&\\ 
    $\rm{Br\beta}$ & 2.6260 & Q(7) 	& 2.4999 &&\\ 
    $\rm{[MgVII]}$ & 3.0276 & Q(8) & 2.5280 &&\\ 
    $\rm{[MgIV]}$ & 4.4870 & Q(9) & 2.5599 &&\\
    && O(3) & 2.8025 && \\ 
    && O(4) & 3.0039 && \\ 
    && O(5) & 3.2349 &&\\ 
    && O(6) & 3.5008 &&\\ 
    && O(7) & 3.8074 &&\rule[-0.9ex]{0pt}{0pt} \\
    \hline
    \end{tabular}
    \caption{This table reports the vacuum wavelengths of the lines marked in \cref{fig:gather}. The predicted wavelengths of the H$_2$ lines are taken from Table 2 in \cite{roueff_full_2019}}.
\end{threeparttable}
\end{table*}

\begin{figure*}[tbp]
    \centering
    \includegraphics[clip, trim=0.25cm 0.25cm 0.25cm 0.25cm, width=0.31\linewidth]{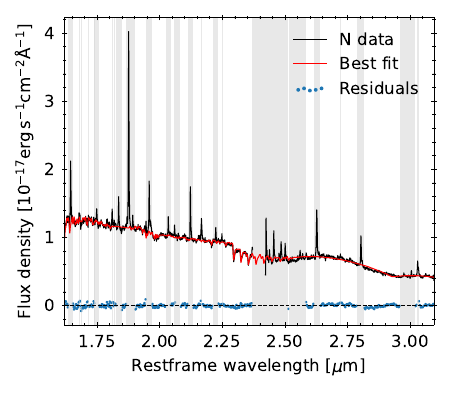}\quad
    \includegraphics[clip, trim=0.25cm 0.25cm 0.25cm 0.25cm, width=0.31\linewidth]{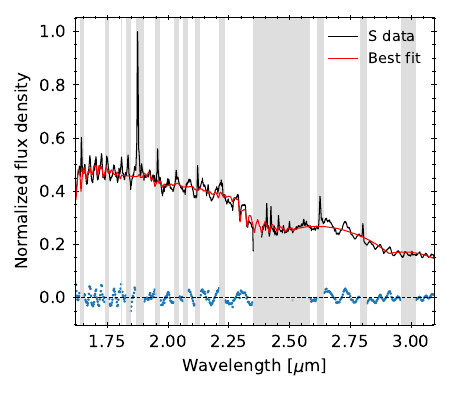}\quad
    \includegraphics[clip, trim=0.25cm 0.25cm 0.25cm 0.25cm, width=0.31\linewidth]{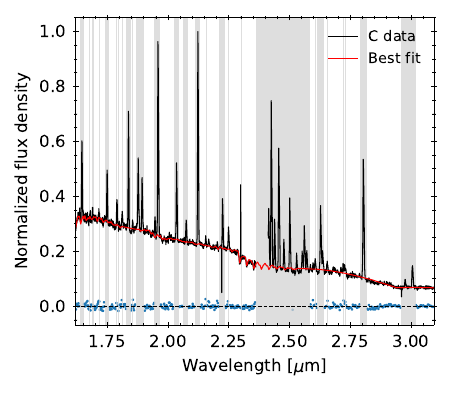}\\
    \includegraphics[clip, trim=0.25cm 0.25cm 0.25cm 0.25cm, width=0.31\linewidth]{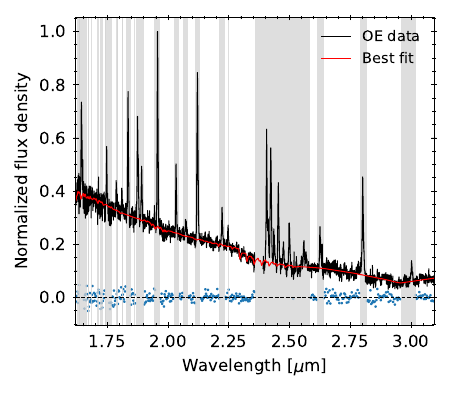}\quad
    \includegraphics[clip, trim=0.25cm 0.25cm 0.25cm 0.25cm, width=0.31\linewidth]{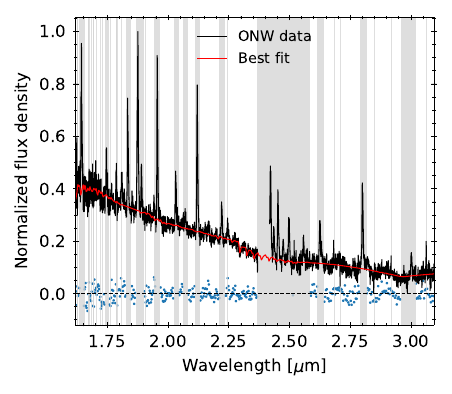}\quad
    \includegraphics[clip, trim=0.25cm 0.25cm 0.25cm 0.25cm, width=0.31\linewidth]{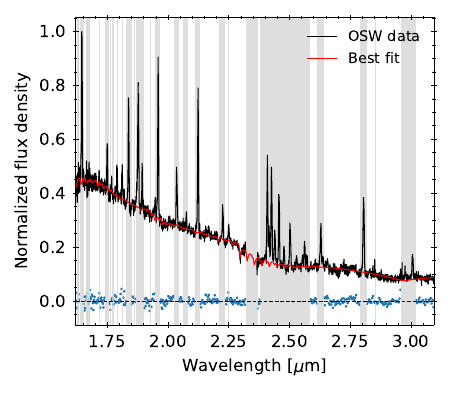}\\
    \caption{Data (black) and best-fit stellar continuum (red) obtained with pPXF in the 1.7~$\mu$m band, for spectra extracted from the individual spaxels marked in \cref{fig:stellar_kin}. The grey shaded regions are masked from the stellar fit.}
    \label{fig:ppxf-results}
\end{figure*}

\begin{figure}[tbp]
    \centering
    \includegraphics[clip, trim=0.25cm 0.25cm 0.25cm 0.25cm, width=0.48\linewidth]{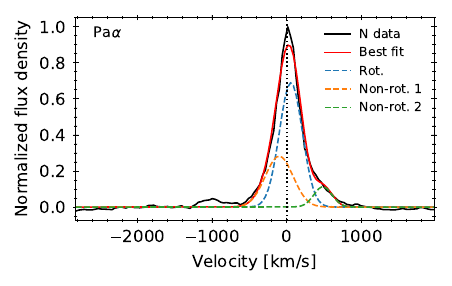}\quad 
    \includegraphics[clip, trim=0.25cm 0.25cm 0.25cm 0.25cm, width=0.48\linewidth]{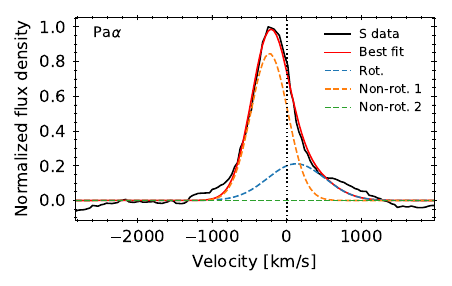}\\ 
    \includegraphics[clip, trim=0.25cm 0.25cm 0.25cm 0.25cm, width=0.48\linewidth]{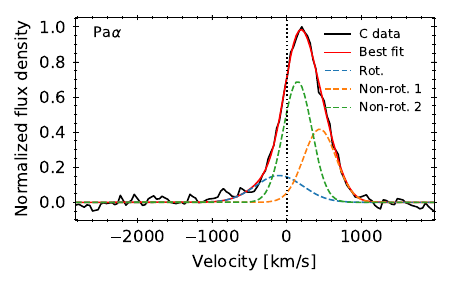}\quad 
    \includegraphics[clip, trim=0.25cm 0.25cm 0.25cm 0.25cm, width=0.48\linewidth]{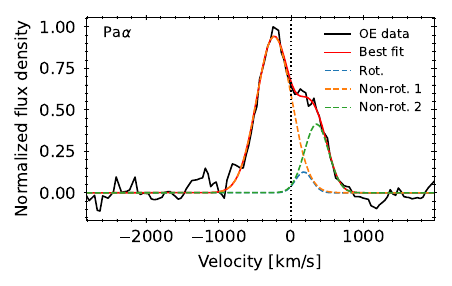}\\ 
    \includegraphics[clip, trim=0.25cm 0.25cm 0.25cm 0.25cm, width=0.48\linewidth]{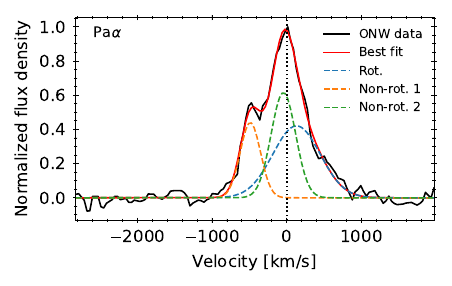}\quad 
    \includegraphics[clip, trim=0.25cm 0.25cm 0.25cm 0.25cm, width=0.48\linewidth]{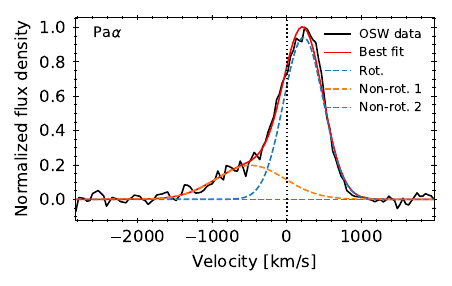}\\ 
    \caption{$\rm{Pa\alpha}$ line fit results, for the six spaxels marked in \cref{fig:stellar_kin}. The observed spectra are shown as black solid curves, and the overall best-fit models as red solid curves. The rotating component is indicated by a blue dashed curve, while the two non-rotating components are shown in orange and green dashed curves, respectively. The spectra have been peak-normalized.}
    \label{fig:pa-fit}
\end{figure}

\begin{figure}[tbp]
    \centering
    \includegraphics[clip, trim=0.25cm 0.25cm 0.25cm 0.25cm, width=0.48\linewidth]{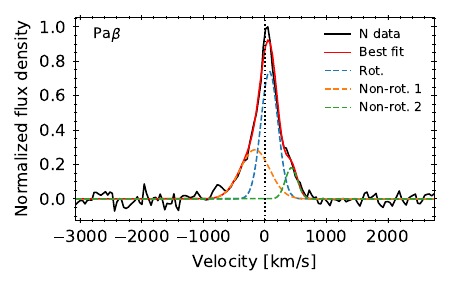}\quad 
    \includegraphics[clip, trim=0.25cm 0.25cm 0.25cm 0.25cm, width=0.48\linewidth]{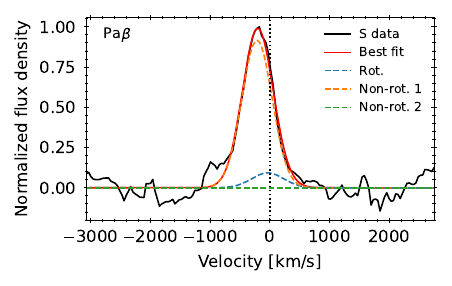}\\ 
    \includegraphics[clip, trim=0.25cm 0.25cm 0.25cm 0.25cm, width=0.48\linewidth]{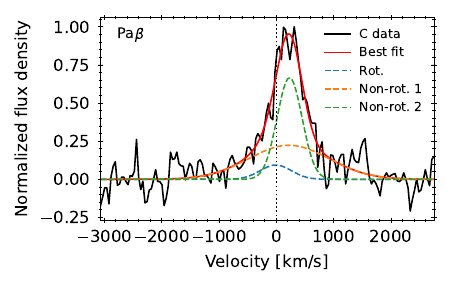}\quad 
    \includegraphics[clip, trim=0.25cm 0.25cm 0.25cm 0.25cm, width=0.48\linewidth]{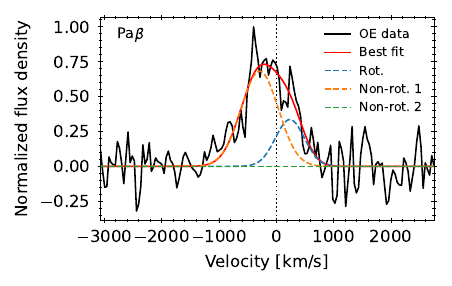}\\ 
    \includegraphics[clip, trim=0.25cm 0.25cm 0.25cm 0.25cm, width=0.48\linewidth]{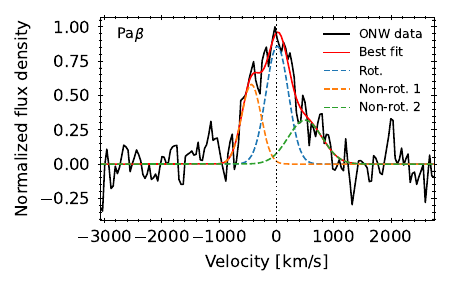}\quad 
    \includegraphics[clip, trim=0.25cm 0.25cm 0.25cm 0.25cm, width=0.48\linewidth]{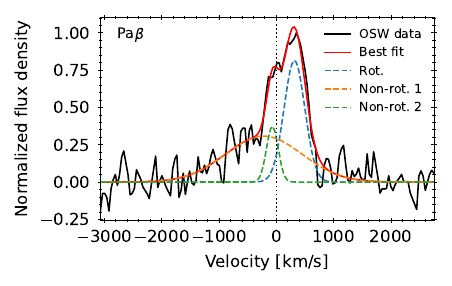}\\ 
    \caption{Same as \cref{fig:pa-fit}, but for the Pa$\beta$ line.}
    \label{fig:pb-fit}
\end{figure}

\begin{figure}[tbp]
    \centering
    \includegraphics[clip, trim=0.25cm 0.25cm 0.25cm 0.25cm, width=0.48\linewidth]{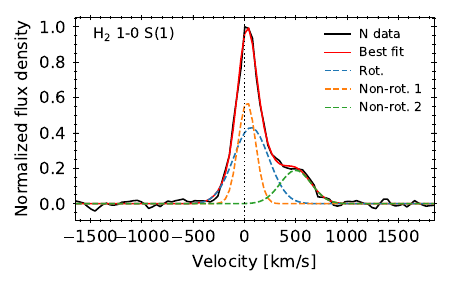}\quad 
    \includegraphics[clip, trim=0.25cm 0.25cm 0.25cm 0.25cm, width=0.48\linewidth]{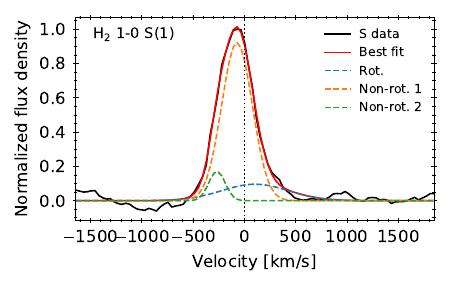}\\ 
    \includegraphics[clip, trim=0.25cm 0.25cm 0.25cm 0.25cm, width=0.48\linewidth]{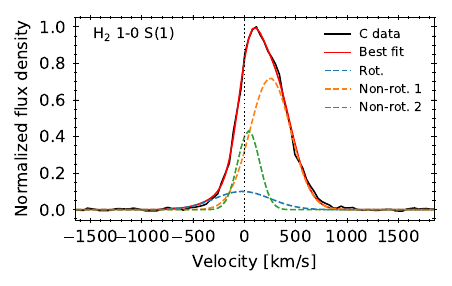}\quad 
    \includegraphics[clip, trim=0.25cm 0.25cm 0.25cm 0.25cm, width=0.48\linewidth]{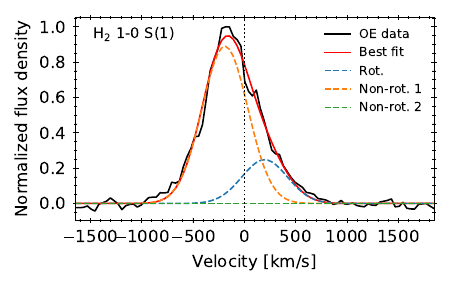}\\ 
    \includegraphics[clip, trim=0.25cm 0.25cm 0.25cm 0.25cm, width=0.48\linewidth]{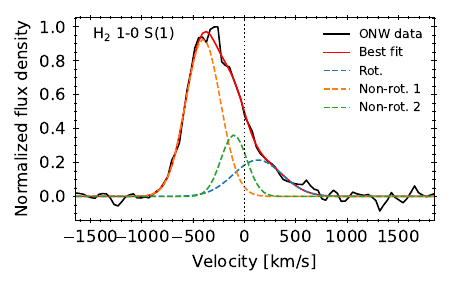}\quad 
    \includegraphics[clip, trim=0.25cm 0.25cm 0.25cm 0.25cm, width=0.48\linewidth]{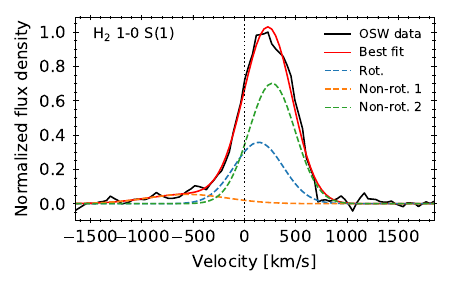}\\ 
    \caption{Same as \cref{fig:pa-fit}, but for the H$_2$ 1-0 S(1) line.}
    \label{fig:h2-fit}
\end{figure}

\begin{figure}[tbp]
    \centering
    \includegraphics[clip, trim=0.25cm 0.25cm 0.25cm 0.25cm, width=0.48\linewidth]{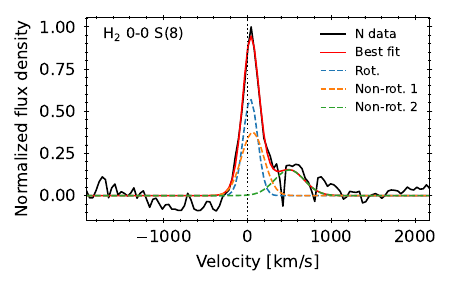}\quad 
    \includegraphics[clip, trim=0.25cm 0.25cm 0.25cm 0.25cm, width=0.48\linewidth]{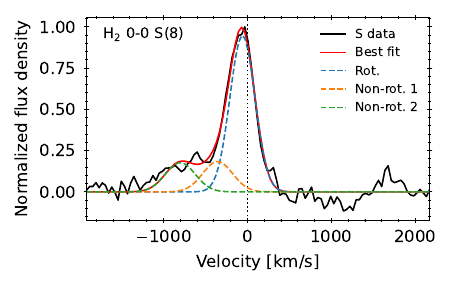}\\ 
    \includegraphics[clip, trim=0.25cm 0.25cm 0.25cm 0.25cm, width=0.48\linewidth]{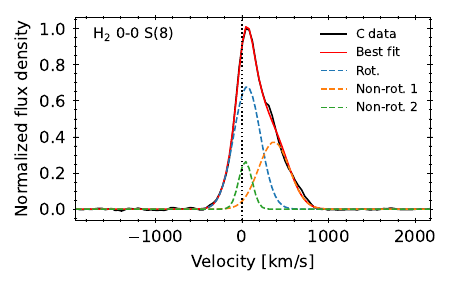}\quad 
    \includegraphics[clip, trim=0.25cm 0.25cm 0.25cm 0.25cm, width=0.48\linewidth]{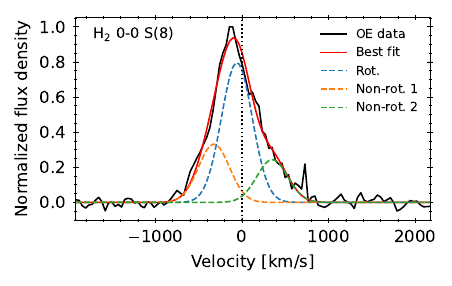}\\ 
    \includegraphics[clip, trim=0.25cm 0.25cm 0.25cm 0.25cm, width=0.48\linewidth]{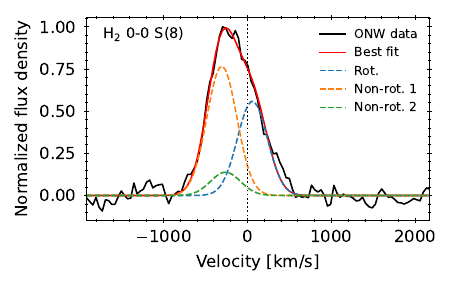}\quad 
    \includegraphics[clip, trim=0.25cm 0.25cm 0.25cm 0.25cm, width=0.48\linewidth]{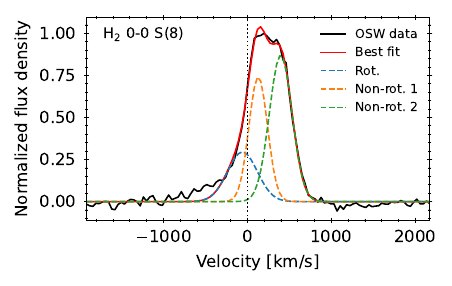}\\ 
    \caption{Same as \cref{fig:pa-fit}, but for the H$_2$ 0-0 S(8) line.}
    \label{fig:h2s8-fit}
\end{figure}

\begin{figure}[tbp]
    \centering
    \includegraphics[clip, trim=0.25cm 0.25cm 0.25cm 0.25cm, width=0.48\linewidth]{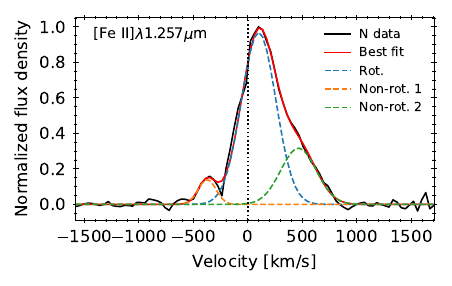}\quad 
    \includegraphics[clip, trim=0.25cm 0.25cm 0.25cm 0.25cm, width=0.48\linewidth]{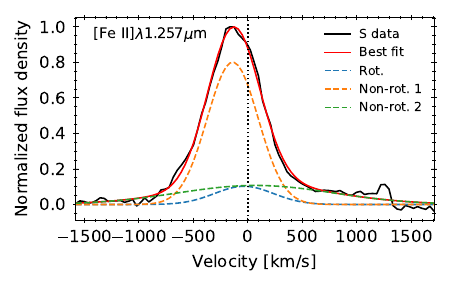}\\ 
    \includegraphics[clip, trim=0.25cm 0.25cm 0.25cm 0.25cm, width=0.48\linewidth]{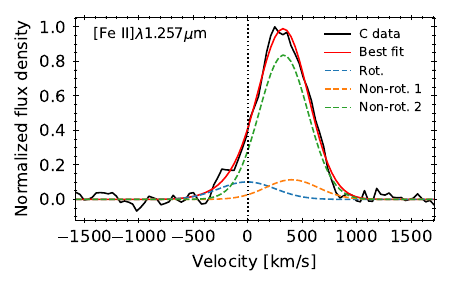}\quad 
    \includegraphics[clip, trim=0.25cm 0.25cm 0.25cm 0.25cm, width=0.48\linewidth]{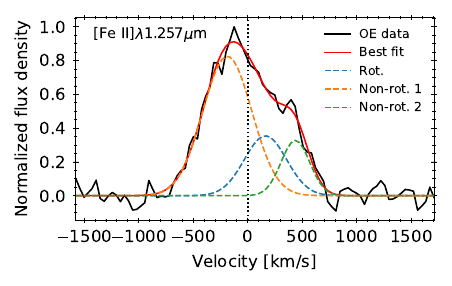}\\ 
    \includegraphics[clip, trim=0.25cm 0.25cm 0.25cm 0.25cm, width=0.48\linewidth]{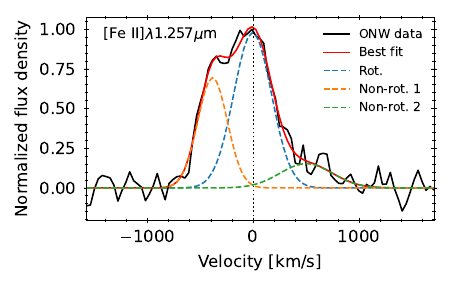}\quad 
    \includegraphics[clip, trim=0.25cm 0.25cm 0.25cm 0.25cm, width=0.48\linewidth]{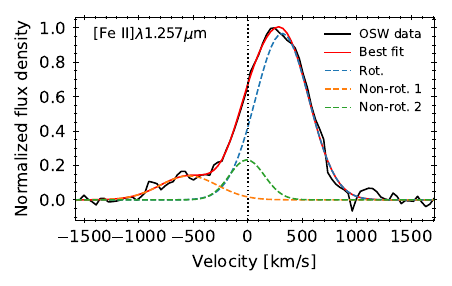}\\ 
    \caption{Same as \cref{fig:pa-fit}, but for the $\rm{[Fe~II]\lambda1257}$ line.}
    \label{fig:fe1257-fit}
\end{figure}

\begin{figure}[tbp]
    \centering
    \includegraphics[clip, trim=0.25cm 0.25cm 0.25cm 0.25cm, width=0.48\linewidth]{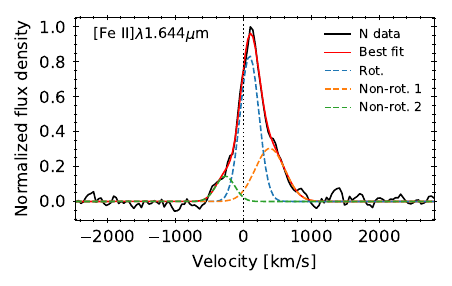}\quad 
    \includegraphics[clip, trim=0.25cm 0.25cm 0.25cm 0.25cm, width=0.48\linewidth]{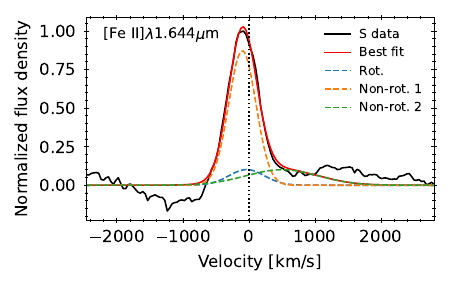}\\ 
    \includegraphics[clip, trim=0.25cm 0.25cm 0.25cm 0.25cm, width=0.48\linewidth]{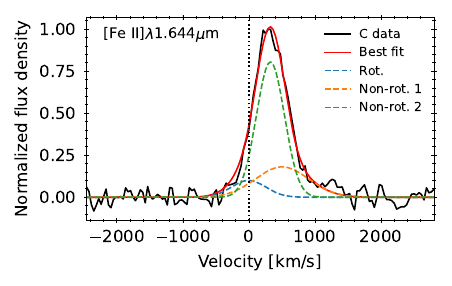}\quad 
    \includegraphics[clip, trim=0.25cm 0.25cm 0.25cm 0.25cm, width=0.48\linewidth]{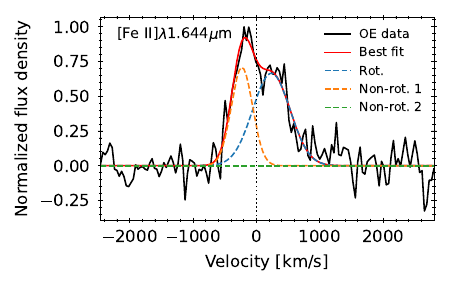}\\ 
    \includegraphics[clip, trim=0.25cm 0.25cm 0.25cm 0.25cm, width=0.48\linewidth]{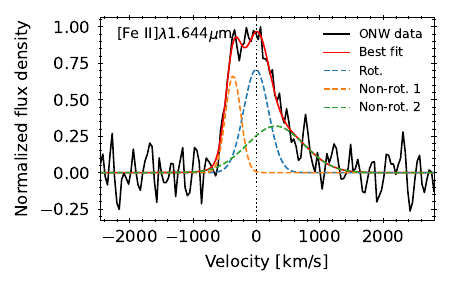}\quad 
    \includegraphics[clip, trim=0.25cm 0.25cm 0.25cm 0.25cm, width=0.48\linewidth]{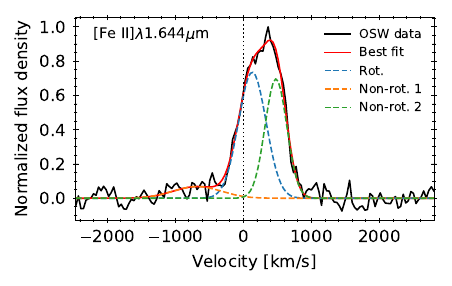}\\ 
    \caption{Same as \cref{fig:pa-fit}, but for the $\rm{[Fe~II]\lambda1644}$ line.}
    \label{fig:fe-fit}
\end{figure}

\begin{figure}[tbp]
    \centering
    \includegraphics[clip, trim=0.25cm 0.25cm 0.25cm 0.25cm, width=0.48\columnwidth]{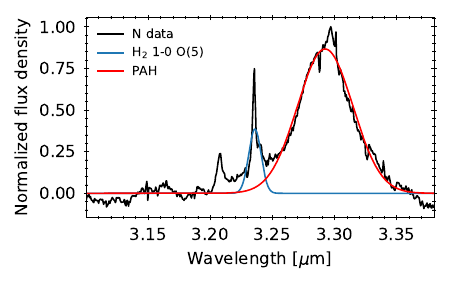}\quad 
    \includegraphics[clip, trim=0.25cm 0.25cm 0.25cm 0.25cm, width=0.48\columnwidth]{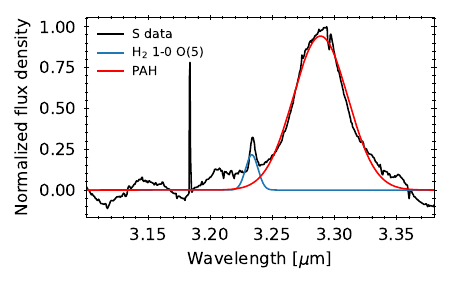}\quad
    \includegraphics[clip, trim=0.25cm 0.25cm 0.25cm 0.25cm, width=0.48\columnwidth]{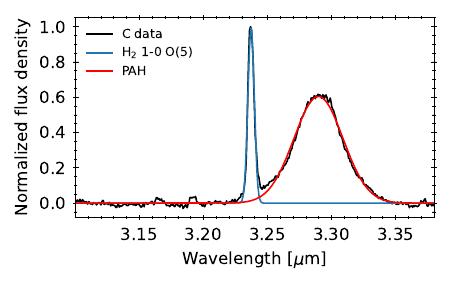}\quad 
    \includegraphics[clip, trim=0.25cm 0.25cm 0.25cm 0.25cm, width=0.48\columnwidth]{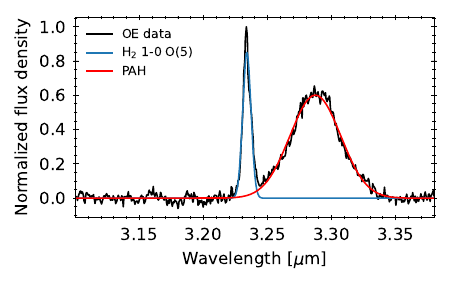}\quad
    \includegraphics[clip, trim=0.25cm 0.25cm 0.25cm 0.25cm, width=0.48\columnwidth]{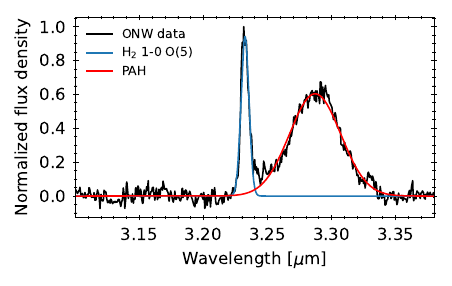}\quad 
    \includegraphics[clip, trim=0.25cm 0.25cm 0.25cm 0.25cm, width=0.48\columnwidth]{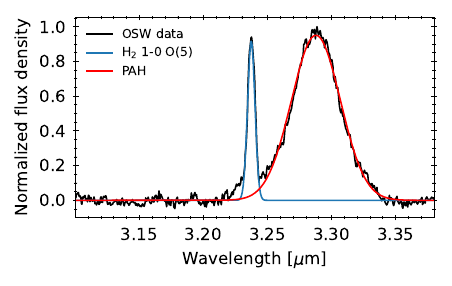}\\ 
    \caption{PAH 3.3 $\mu$m fit results for the six spaxels marked in \cref{fig:stellar_kin}. The blue curve is the fit to the $\rm{H_2\,1-0\,O(5)}$ emission line, which is detected in all spaxels.}
    \label{fig:pah-spaxels-fits}
\end{figure}

\end{document}